\documentclass[referee]{raa}           
\usepackage{tikz}
\usepackage{graphicx,times}
\usepackage{natbib}
\usepackage{amssymb,amsmath}
\bibpunct{(}{)}{;}{a}{}{,}

\usepackage[pagebackref=true]{hyperref}

\begin{document}

   \title{Development of 6-inch 80-170~GHz broadband silicon plated horn antenna arrays for primordial gravitational wave search}

 \volnopage{ {\bf 2024} Vol.\ {\bf X} No. {\bf XX}, 000--000}
   \setcounter{page}{1}

\author{
Yuan-hang He\inst{1,2}, 
Shi-bo Shu\inst{2,3}\thanks{E-mail: shusb@ihep.ac.cn}, 
Ya-qiong Li\inst{2,3}, 
Xue-feng Lu\inst{2,3}, 
Ye Chai\inst{1,2}, 
Xiang Li\inst{4},
\\Zhi Chang\inst{2}, 
He Gao\inst{2},
Yu-dong Gu\inst{2},
Xu-fang Li\inst{2},
Zheng-wei Li\inst{2,3},
Zhou-hui Liu\inst{2},
\\Guo-feng Wang\inst{2},
Zhong-xue Xin\inst{2},
Dai-kang Yan\inst{2,3},
Ai-mei Zhang\inst{2,3},
Yi-fei Zhang\inst{2,3},
\\Yong-jie Zhang\inst{2,3},
Wen-hua Shi\inst{3,4},
Jue-xian Cao\inst{2,3}\thanks{E-mail: jxcao@xtu.edu.cn}, 
Cong-zhan Liu\inst{2,3},
}

\institute{ School of Physics and Photoelectric Engineering, Xiangtan University, Xiangtan 100871, China\\
\and
Key Laboratory of Particle Astrophysics,Institute of High Energy Physics, Chinese Academy of Sciences, Beijing 100049, China\\
\and
University of Chinese Academy of Sciences, The University of Science
and Technology of China, Chinese Academy of Sciences, Hefei, Anhui, 230026, China\\
\and 
Suzhou Institute of Nano-Tech and Nano-Bionics, Chinese Academy of Sciences, Suzhou 215123, China\\
\vs \no
   {\small Received 20XX Month Day; accepted 20XX Month Day}
}

\abstract{Searching for primordial gravitational wave in cosmic microwave background (CMB) polarization signal is one of the key topics in modern cosmology. Cutting-edge CMB telescopes requires thousands of pixels to maximize mapping speed. Using modular design, the telescope focal plane is simplified as several detector modules. Each module has hundreds of pixels including antenna arrays, detector arrays, and readout arrays. The antenna arrays, as the beam defining component, determine the overall optical response of the detector module. In this article, we present the developments of 6-inch broadband antenna arrays from 80~GHz to 170~GHz for the future IHEP focal plane module. The arrays are fabricated from 42 6-inch silicon wafers including 456 antennas, 7~\% more pixels than usual design. The overall in-band cross polarization is smaller than -20 dB and the in-band beam asymmetry is smaller than 10\%, fulfilling the requirements for primordial gravitational wave search.
}

\authorrunning{Y.-H. He et al. }            
\titlerunning{Horn antenna arrays for primordial gravitational wave search}  
\maketitle

%
\section{Introduction}           
\label{sect:intro}
The observations of CMB hold immense significance in cosmology research. This ancient light, leftover from the Big Bang, provides crucial insights into the early universe’s conditions, such as its temperature, density, and composition~(\cite{abitbol2017cmb}). CMB mainly has two polarization modes: E-mode and B-mode, analog to the electromagnetic field~(\cite{britton2009progress, hu1997cmb}). B-mode polarization could be generated by primordial gravitational waves, which could provide crucial information on the inflation process. Therefore, searching for the large-scale CMB B-mode polarization signals has been the next goal in observational cosmology(~\cite{campeti2024litebird}). 

For ground-based CMB observations, the atmosphere poses a significant challenge. It absorbs CMB signal, and emits radiation in the same frequency range, which degrades the signal-to-noise ratio. Among all atmospheric components, water vapor plays a crucial role due to its strong absorption. Therefore, CMB telescopes are located at dry sites in the world. For example, the South Pole Telescope(~\cite{anderson2018spt}) and the BICEP telescopes(~\cite{ade2016bicep2}) are located in Antarctica, the Atacama Cosmology Telescope(~\cite{hincks2010atacama}) and the Simons Observatory are located in Chile(~\cite{arnold2010polarbear}), and the Ali CMB Polarization Telescope (AliCPT), of which the first telescope is under deployment, is located at the Qinghai-Tibet Plateau in Xizang, China(~\cite{salatino2020design, li2019probing}). All these telescopes use modular focal plane design. The cutting-edge detector module is based on the 6-inch micro-fabrication process(~\cite{anderson2018spt}). Each detector module consists of three main parts: antenna arrays, superconducting detector arrays, and readout arrays. As the signal receiving component, the antenna arrays design is crucial for the overall observation performance. 

Horn antennas are widely used, because of its excellent polarization definition and broadband frequency coverage(~\cite{wang2017wideband}). There are mainly two kinds of horn antennas used in B-mode detection: corrugated horns and smooth-walled horns. Corrugated horns have near ideal beam symmetry, but the loaded rings are larger than the aperture , decreasing the number of pixels in a given module size. Therefore the mapping speed is not optimal. Smooth-walled horns with no extra area wasted can achieve the optimal number of pixels(~\cite{simon2016design}). We will present our single horn design and the antenna arrays design.

Fabrication of horn arrays is related to many aspects of the detector module assembly. Direct-machining has the advantages of low cost and easy fabrication, but the machining precision and the discrepancy of coefficients of thermal contraction between metal and silicon requires special cares. Micro-fabrication technology provides a fabrication precision of a level of several micrometers. The antenna arrays and the detector arrays are both fabricated from silicon wafers, providing no thermal relative contraction at cryogenic temperature(~\cite{thornton2016atacama}). 

In this paper, we present the design, fabrication, and measurements of a 6-inch silicon-plated smooth-walled horn arrays. The frequency range is from 80~GHz to 170~GHz covering the two main atmospheric windows for the CMB observations. AliCPT-1(~\cite{salatino2020design}) is the first CMB telescope of the AliCPT project leading by Institute of High Energy Physics (IHEP), Chinese Academy of Sciences, and also the first CMB telescope in China. Its focal plane can hold 19 detector modules. For now, there is one detector module deployed in the current phase, fabricated by Stanford University and National Institute of Standards and Technology(~\cite{salatino2021current}). This new antenna arrays design presented in this paper, has a total number of 456 pixels, $7\%$ more than the current AliCPT-1 module. This antenna arrays could be used for future AliCPT telescopes.

\section{Silicon-plated horn antenna design}
\label{sec:design}

\subsection{Antenna profile definition}
\label{sec:ant_def}

To match the starting frequency of 80~GHz, an input radius $r_{\text{in}}$ = 1.15~mm is used with a cutoff frequency $f_{\textrm{cut-off}}=76.4$~GHz. The horn diameter is determined by the optimization of the observing speed. In general, the aperture efficiency is determined by the spillover efficiency and the edge taper. The maximum aperture efficiency happens when the horn diameter $2r_{\text{out}}$=2F$\lambda$, where F is the f-number of the telescope, and $\lambda$ is the wavelength(~\cite{griffin2002relative}). Here, we take the optics parameters of the AliCPT-1 telescope(~\cite{salatino2020design, salatino2023laboratory}) as the inputs for our antenna design. AliCPT-1 has F=1.4 and $\lambda=$ 2.4~mm in the center of the 80-170~GHz observation band. The horn diameter is calculated to be $2r_{\text{out}}$=2F$\lambda$=6.72~mm for maximum aperture efficiency. However, a maximum aperture efficiency does not guarantee a maximum observing speed. Given a fixed focal plane size, more pixels are preferred in CMB observations. In our case, 1.54~F$\lambda$ is designed as the horn diameter with $r_{\text{out}}$ = 2.6~mm. 

The horn antenna profile is defined by a monotonically increasing curve given by Simon et al.(~\cite{simon2016cosmic}). This curve provides the flexibility of producing a continuous multi-section smooth-walled profile with only 42 parameters. The horn profile is simulated in the CORRUG software, an antenna simulation software based on mode-matching method(~\cite{mitra1963relative, polo2017analysis}). The curve step is 0.25~mm, as two different thickness (0.25~mm and 0.50~mm) of 6-inch wafer are used for practical reason. The initial profile has a length of 19.5~mm with 78 sections for further optimization.

\subsection{Antenna profile optimization}
\label{sec:ant_opt}

The beam asymmetry will introduce extra difficulty for polarization calibration. Also it is hard to obtain a high symmetry antenna within such a broadband from 80~GHz to 170~GHz, so the primary optimization goal is to minimize the overall in-band beam asymmetry. The penalty function $P$ is defined as
\begin{equation}
    P\equiv \sum_{\text {Frequency }} \sum_{\theta=0}^{\theta=\theta_{\text {stop }}}\left |E^{2}-H^{2}\right |,
\end{equation}
where frequency range is calculated from 80~GHz to 170~GHz. The cold stop has cut-off at $\theta_{\text {stop}}=18.5^\circ$. Markov Chain Monte Carlo (MCMC) method is used to randomly create parameters based on the current best $P$. When a smaller $P$ was achieved, this process was repeated again.

The initial values of the 42 parameters are randomly given. Three parameters are fixed: $r_{\text{in}}$ = 1.15~mm, $r_{\text{out}}$ = 2.6~mm, and the total length of 19.5~mm. The calculated horn profile with a resolution of 0.25~mm is simulated in CORRUG. Far-field beam patterns are calculated with a frequency resolution of 1~GHz and an angle resolution of 1~degree. $P$ is calculated from the beam patterns. In general a monotonically increasing profile usually provides a broadband low return loss S11, so we only check S11 at the end of full optimization.

\begin{figure}[h]
   \centering
   \begin{subfigure}{0.45\textwidth}
        \centering
        \includegraphics[width=\linewidth]{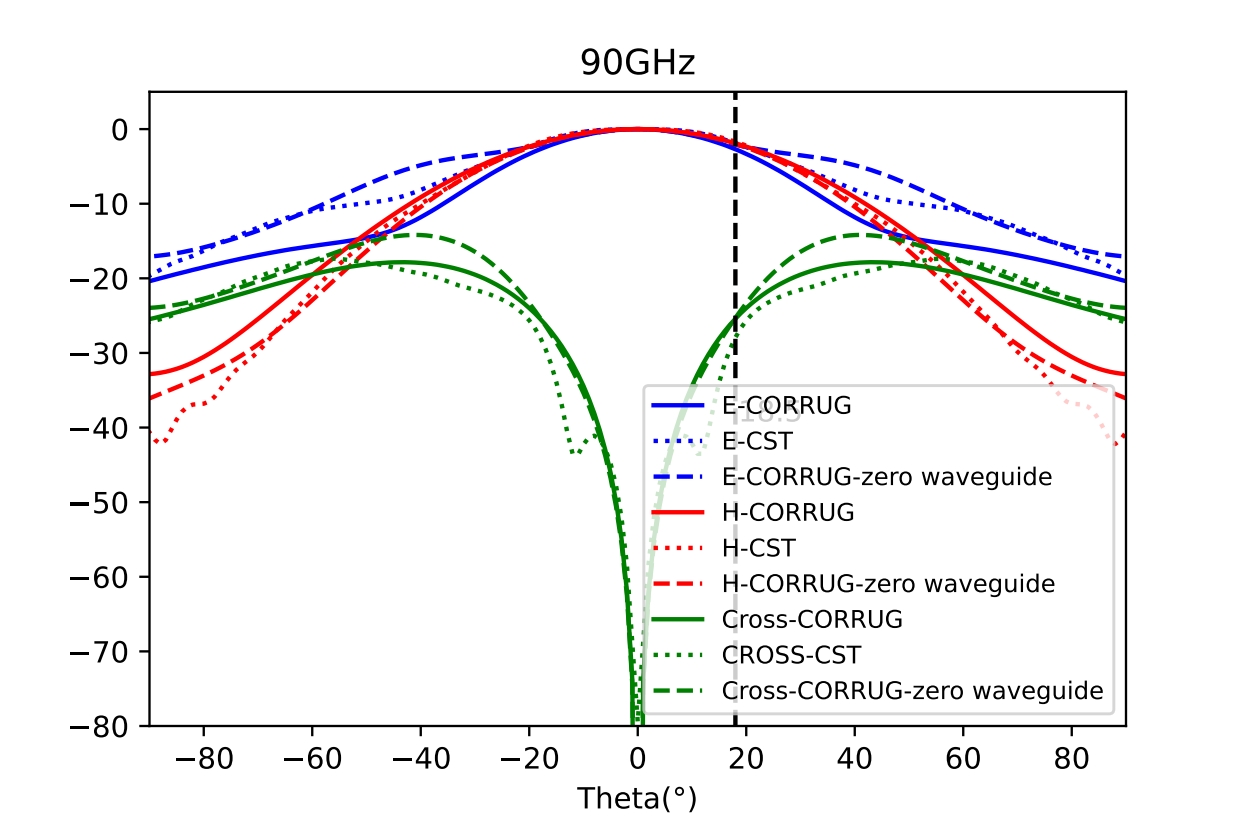}
    \end{subfigure}
    \hspace{0.05\textwidth}
    \begin{subfigure}{0.45\textwidth}
        \centering
        \includegraphics[width=\linewidth]{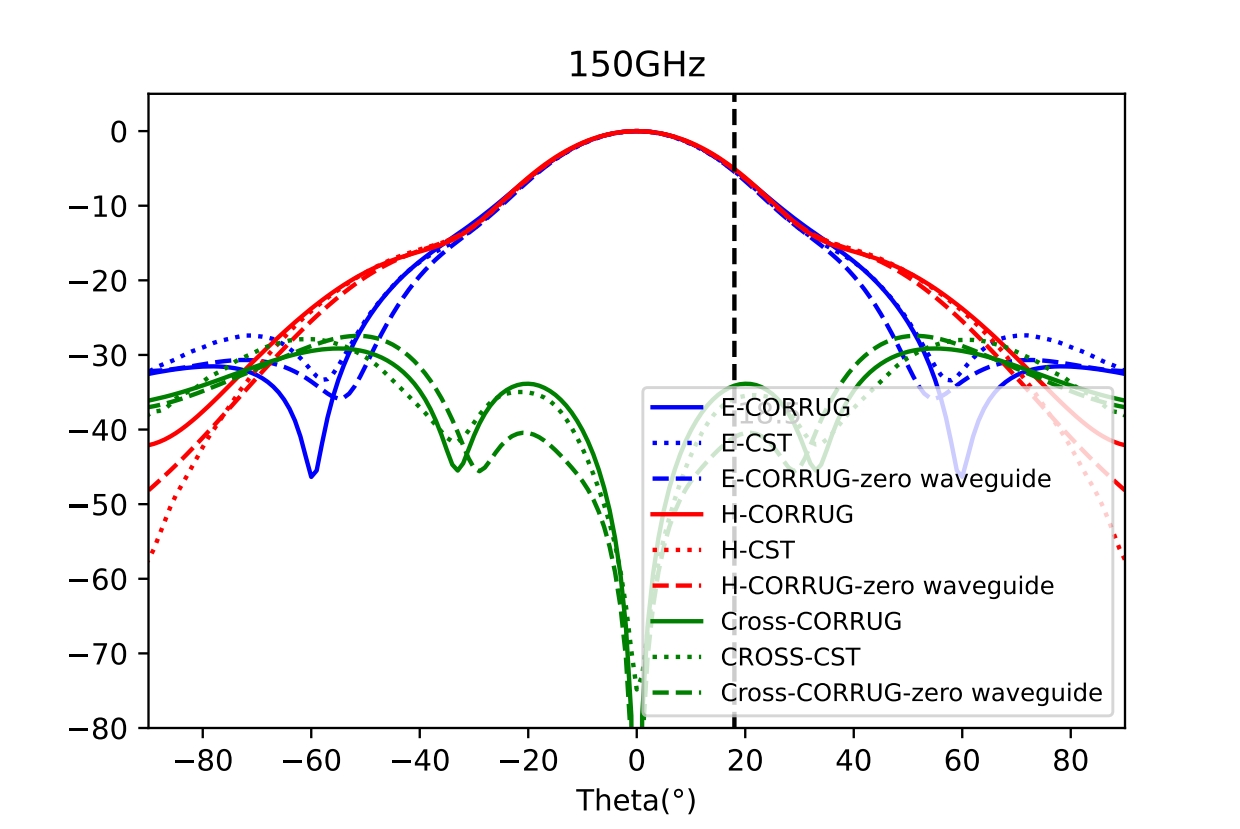}
    \end{subfigure}
   \caption{The solid line represents the beam pattern of the antenna simulated by CORRUG without the zero-length waveguide section. The two different dashed lines represent the beam patterns of the antenna simulated by CORRUG with the zero-length waveguide section and the beam pattern simulated by CST, respectively. The CST simulation uses an antenna model that is consistent with the actual fabricated antenna.}
    \label{fig:zero waveguide}
\end{figure}

Compared with 3D finite element simulation softwares like Ansys HFSS and CST, the software using mode-matching method, like the CORRUG software, in general provides much faster simulations. In our case, the far-field beam calculation in CORRUG only costs 14 seconds, while it costs about 30 minutes in CST. We need tens of thousands of optimizations for this 42-parameter design, so CORRUG is selected. CORRUG software uses mode-matching calculations in each waveguide section, but an assumption that the aperture is well matched to the free-space is made. This assumption works well when the aperture diameter is 2.5 times larger than the wavelength according to the software manual. In our case, with $r_{\text{out}}$ = 2.6~mm this assumption only works when the frequency is higher than 144~GHz. According to the CORRUG manual, adding an extra large-diameter waveguide section with zero thickness in calculation may provide more accurate results. We compared the CORRUG results by add an extra zero-thick waveguide in Fig.~\ref{fig:zero waveguide}. Here we use the CST results as the standard, which is valid in our measurements in Sec.~\ref{sec:single_meas}. Compared with CST results, The CORRUG gives similar results within 20~° especially at high frequencies. The discrepancy begins noticeable when the angle is larger than 20~°. At 90~GHz, CORRUG with extra waveguide gives a more accurate H-plane results, but there is a big discrepancy in the E-plane around 20~°. Therefore, no extra waveguide is added in our CORRUG simulations. The discrepancy between simulation and measurements will be discussed in Sec.~\ref{sec:single_meas}.

Only using the asymmetry as the optimization aim will result in a wide main-beam by decreasing the diameter close to the aperture. This means the fixed horn aperture is not fully utilized. In this case, the beam coupling efficiency, given by 
\begin{equation}
    \text { Beam Coupling Efficiency }=\frac{\int_{0}^{\theta_{\text {stop }}} \frac{1}{2}\left(E^{2}+H^{2}\right) \sin \theta d \theta}{\int_{0}^{90^{\circ}} \frac{1}{2}\left(E^{2}+H^{2}\right) \sin \theta d \theta} \,
\end{equation}
will be surprisingly low. To avoid this situation, three criteria are given based on Simon et al(~\cite{simon2016cosmic}) and practical reasons: 1. the profile gradient must not exceed 0.1~mm per section near the aperture; 2. the maximum difference in radius between adjacent layers must not exceed 0.49~mm; 3. The last third of the total antenna profile length must be greater than 0.69 times the aperture diameter but less than 0.88 times the aperture diameter.

\begin{figure}[h]
    \centering
    \begin{subfigure}{0.7\textwidth}
        \centering
        \includegraphics[width=\linewidth]{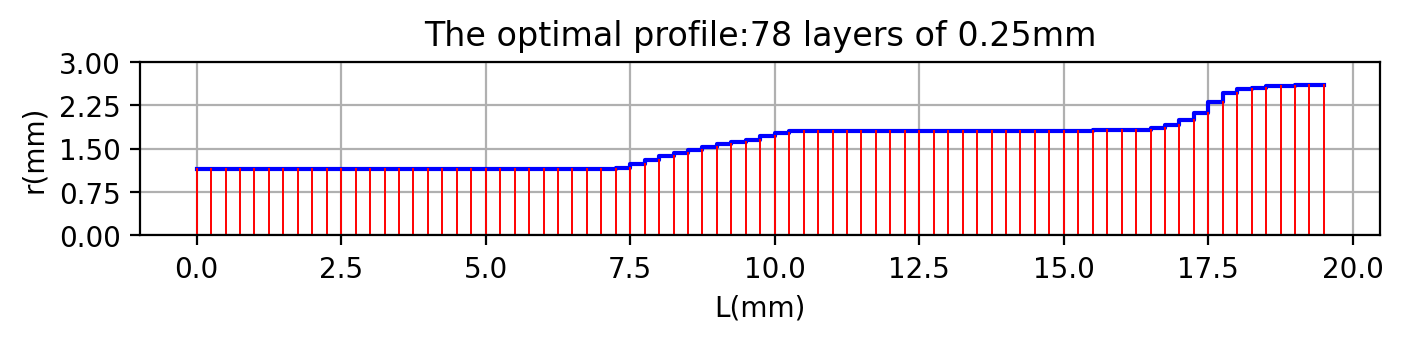}
        \tikz{(a) Optimized antenna profile}
    \end{subfigure}
    \hspace{0.05\textwidth}
    \begin{subfigure}{0.7\textwidth}
        \centering
        \includegraphics[width=\linewidth]{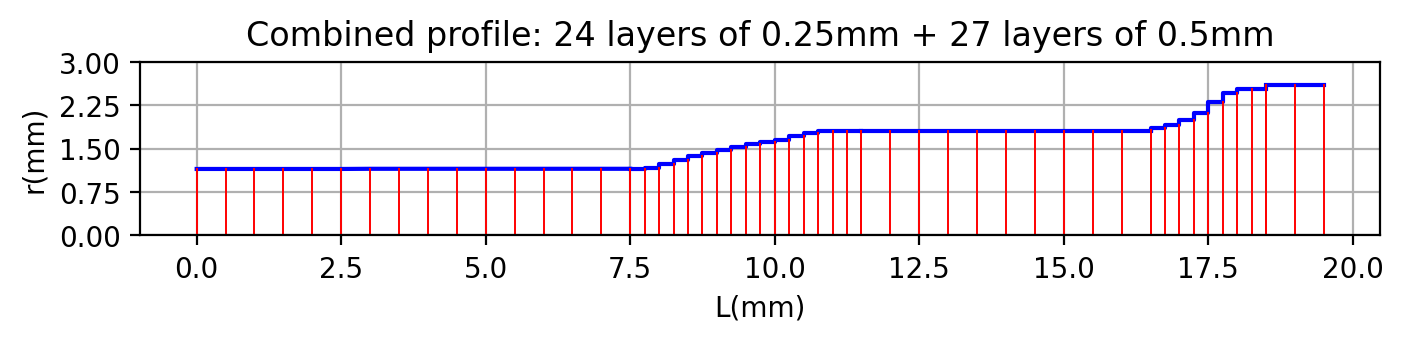}
        \tikz{(b) Combined antenna profile}
    \end{subfigure}
    \caption{(a) The final optimized antenna profile, which consists of 78 layers of 0.25~mm sections. (b) The combined antenna profile obtained from Figure (a), which consists of 24 layers of 0.25~mm sections and 27 layers of 0.25~mm sections.}
    \label{fig:ant profile}
\end{figure}

\begin{figure}[h]
    \centering
    \begin{subfigure}{0.4\textwidth}
        \centering
        \includegraphics[width=\linewidth]{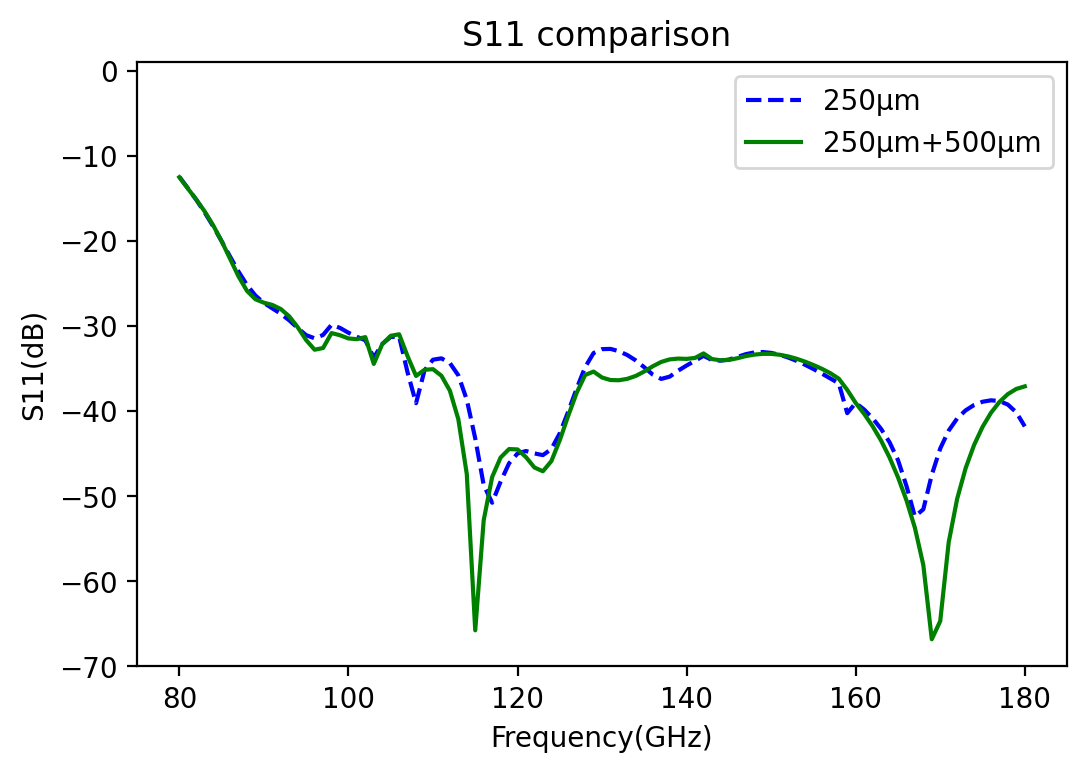}
    \end{subfigure}
    \hspace{0.05\textwidth}
    \begin{subfigure}{0.4\textwidth}
        \centering
        \includegraphics[width=\linewidth]{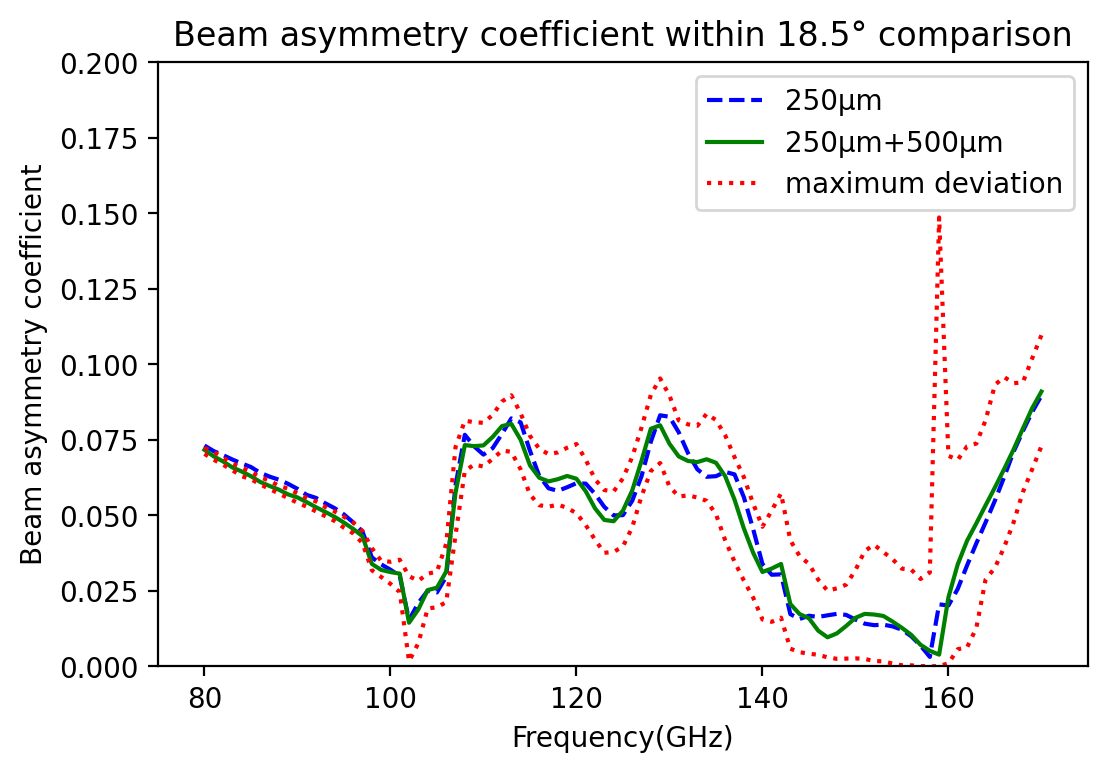}
    \end{subfigure}
    \hspace{0.05\textwidth}
    \begin{subfigure}{0.4\textwidth}
        \centering
        \includegraphics[width=\linewidth]{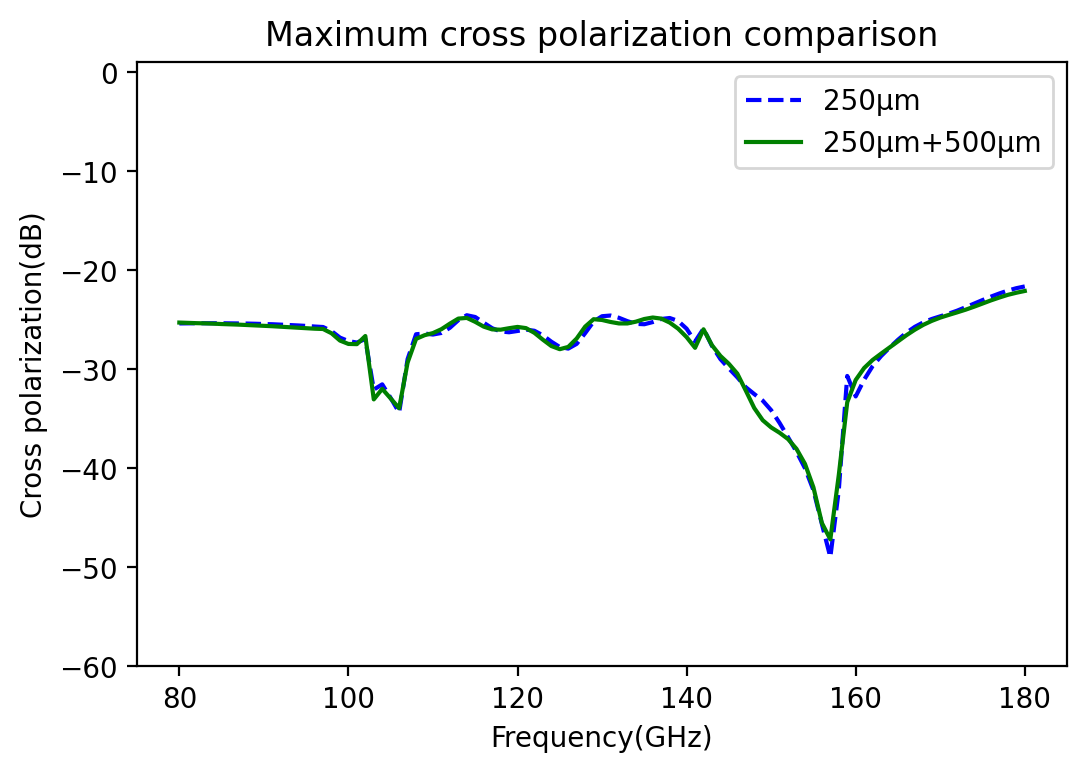}
    \end{subfigure}
    \hspace{0.05\textwidth}
    \begin{subfigure}{0.4\textwidth}
        \centering
        \includegraphics[width=\linewidth]{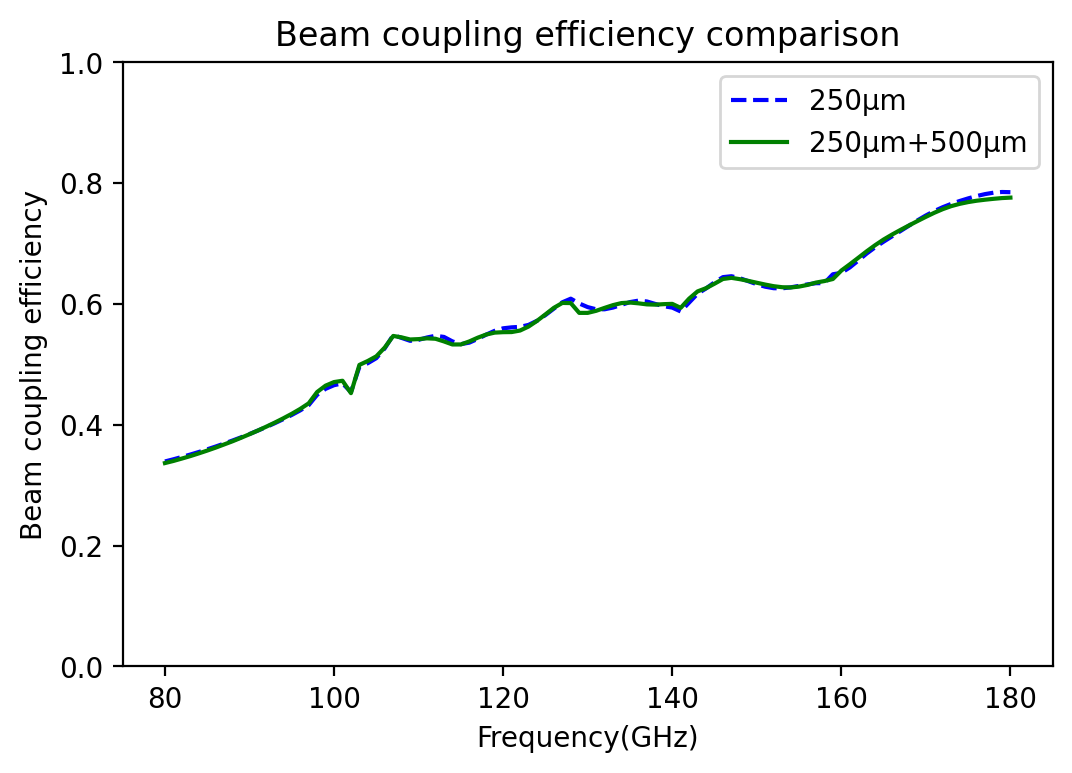}
    \end{subfigure}
    \caption{The dashed line represents the simulation performance of the antenna before merging, which was made from 78 layers of 0.25~mm thick silicon wafers stacked together; the solid line represents the simulation performance of the antenna after merging, which was made from 24 layers of 0.25~mm thick silicon wafers and 27 layers of 0.5~mm thick silicon wafers stacked together. The red dot curves show the maximum deviations in the fabrication error analysis.}
    \label{fig:combine}
\end{figure}

At each frequency, the beam asymmetry coefficient is defined as:
\begin{equation}
    \text { Beam asymmetry coefficient }=\frac{\sum_{\theta=0}^{\theta=\theta_{\text {stop }}}\left | E^{2} -H^{2} \right |}{\sum_{\theta=0}^{\theta=\theta_{\text {stop }}} E^{2}}.
\end{equation}
This equation sums the difference between the radiated power in the E-plane and H-plane over angles from $\theta$=0 to $\theta_{stop}$=18~°. Then, this summation is normalized to the E-plane radiation power. A small beam asymmetry coefficient is preferred.

\subsection{Final design}
\label{sec:final_design}

Initially, five random linear horn antenna profiles are generated. The configuration file for each antenna profile is then sent to CORRUG for simulation and optimization. Each initial profile undergoes a total of 10,000 iterations. Finally, the profile with the lowest penalty function value is selected from the five iteration results as the final antenna design. The final optimization result is shown in Fig.~\ref{fig:ant profile}~(a). We further manually combined the sections with similar radii into 0.5~mm-thick sections to reduce the number of sections from 78 to 51, The combined antenna profile is shown in Fig.~\ref{fig:ant profile}~(b). The total number of 6-inch silicon wafers required to fabricate a single silicon horn antenna is two. One 0.25-mm-thick wafer is patterned into 24 pieces, and one 0.5-mm-thick wafer is patterned into 27 pieces. The S11, beam asymmetry, maximum cross polarization, and beam coupling efficiency were compared in Fig.~\ref{fig:combine}. The difference after the combination is negligible.

To consider the unavoidable error from fabrication, we applied a random error to the radius of each layer in the 51-layer final design. The error has a uniform distribution within $\pm$7~\textmu m, an estimation of fabrication and assembling errors. With 1000 simulations, the floating range of deviation in the antenna beam asymmetry coefficient is below 10\% for most frequency points, shown in Fig.~\ref{fig:combine}, indicating that the errors do not significantly affect the antenna symmetry and are acceptable. There is a large deviation at 159~GHz, due to a specific structure from only one of the 1000 random profiles at that particular frequency point. This error analysis suggests that the antenna performance is quite robust to the geometry errors.

\begin{figure*}[h]
   \centering
   \includegraphics[width=0.95\textwidth]{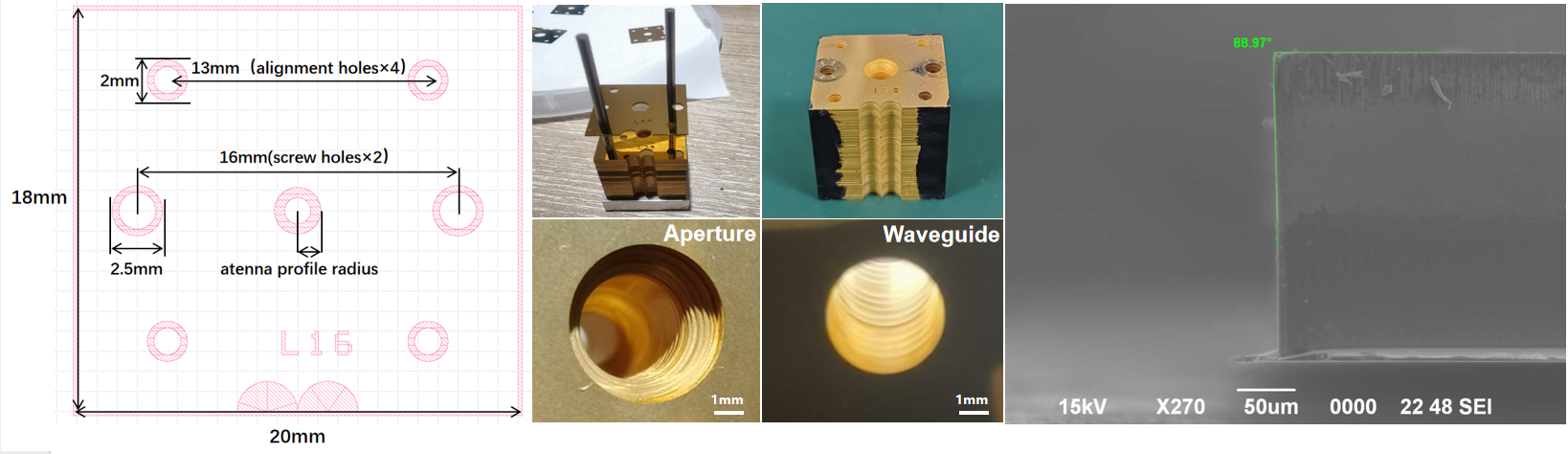}
   \caption{The illustration of the antenna stacking assembly, along with the characterization results of the hole structures in a single-layer silicon wafer and the verticality of deep silicon etching.}
\label{fig:assembly}
\end{figure*}

The design of the mask for single horn fabrication is shown in Fig.~\ref{fig:assembly}. Each piece contains one antenna hole, two alignment holes, and four screw holes for stacking following our previous design(~\cite{feng2022development}). Only circular rings are etched for those holes to decrease the etching area and increase the etching uniformity.

To match standard rectangular waveguides in measurements, two rectangle-to-circle waveguide adapters are designed and machined: a WR10 one for 75-110~GHz and a WR6 one for 110-170~GHz. The design is simply a linear transition structure with a length of 25.4~mm. The S11 of both adapters are less than -30~dB.

\begin{figure}[h]
   \centering
   \includegraphics[width=0.95\textwidth]{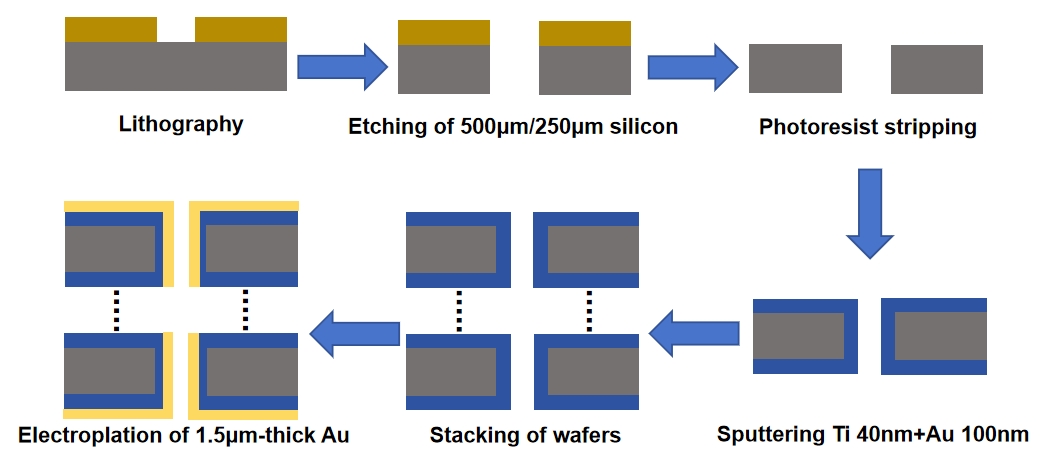}
   \caption{Micro-fabrication process flow diagram for a single silicon-plated smooth-walled horn antenna.}
\label{fig:fab}
\end{figure}

\section{Fabrication of a single silicon-plated horn antenna}
\label{sec:single_fab}

6-inch silicon wafers are used in our single horn fabrication as a pathfinder for the final 6-inch horn arrays fabrication. Contact lithography provides an accuracy of 1~\textmu m, which is much higher than usual metal machining. The whole fabrication process, shown in Fig.~\ref{fig:fab}, was done in the cleanroom of Suzhou Institute of Nano-Tech and Nano-Bionics (SINANO), Chinese Academy of Sciences. After contact lithography, wafers are etched through the Bosch process in a SPTS deep reactive-ion etching tool. The etched sidewall angle of 89~° is achieved at one end, shown in Fig.~\ref{fig:assembly}. Given the wafer thickness of 0.25~mm and 0.5~mm, this angle provides maximum geometry deviations of 4.4~\textmu m and 8.7~\textmu m, respectively. As the slope is very close to 90~° at the other end, the overall deviations would be smaller than this calculation.

After deep silicon etching, small pieces are all released from the wafer. On both sides of those pieces, 40~nm-thick Ti and 100~nm-thick Au are sputtered as a thin seed layer of subsequent gold electroplating. Sputtering is selected for a good step coverage. We stack small silicon pieces on a custom jig using two alignment pins with a tolerance of ±1\textmu m. To finish the assembly, two screws are used as shown in Fig.~\ref{fig:assembly}. Afterwards, this single antenna is electroplated with 1.5~\textmu m-thick gold. Epoxy resin STYCAST 2850FT is used to the sidewalls as a glue to secure the entire antenna assembly and screws are removed.

\section{Measurements of a single horn antenna}
\label{sec:single_meas}

All measurements except the antenna gain were performed in the microwave darkroom at the Key Laboratory of Particle Astrophysics, IHEP. The antenna gain was measured The frequency range is 75-170~GHz to cover the cutoff frequency. Two standard gain antennas from ERAVANT, SAR-2309-10-S2 (23~dBi) and SAZ-2410-06-S1 (24~dBi), are used as the transmitters for frequency ranges of 75-110~GHz and 110-170~GHz, respectively.

\begin{figure}[h]
    \centering
    \begin{subfigure}{0.4\textwidth}
        \centering
        \includegraphics[width=\linewidth]{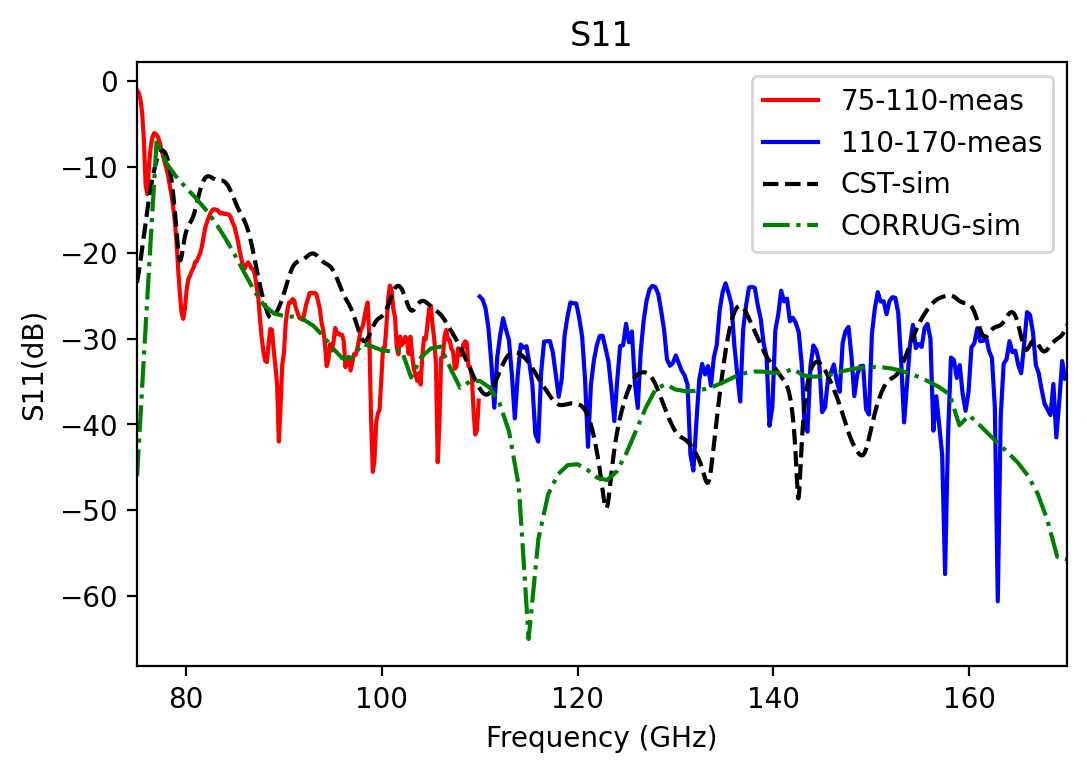}
        \tikz{(a) S11}
    \end{subfigure}
    \hspace{0.05\textwidth}
    \begin{subfigure}{0.4\textwidth}
        \centering
        \includegraphics[width=\linewidth]{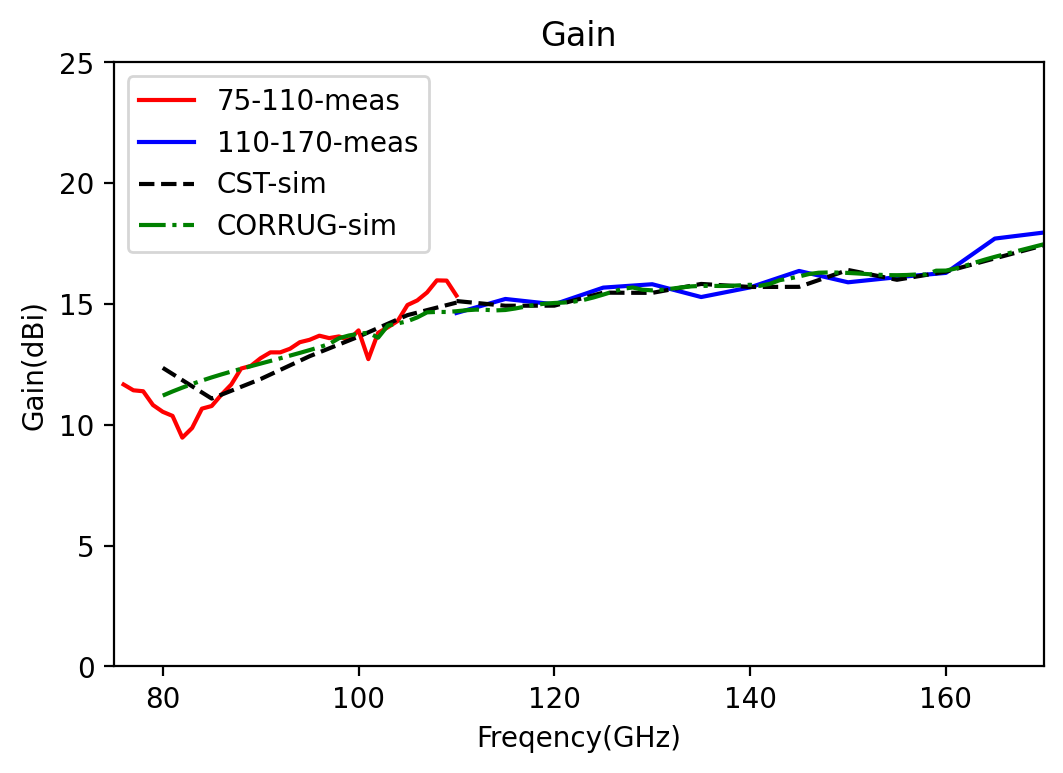}
        \tikz{(b) Gain}
    \end{subfigure}
    \caption{(a): Comparison between measured (solid line) and simulated (dashed line) the AUT’s S11 results. (b): Comparison between measured (solid line) and simulated (dashed line) the AUT’s gain results.}
    \label{fig:S11 and gain}
\end{figure}

In Fig.~\ref{fig:S11 and gain}, the S11 has been calibrated to remove the reflection of the rectangular-to-circular waveguide adapter. The measured S11 agrees well with both simulations, especially the CST result. The ripple at low frequency in CST is captured well. The S11 is below -10~dB from 78~Ghz to 170~GHz as designed. The gain is measured by comparing the AUT with a standard gain antenna. The overall gain also agrees well with simulations. Given a fixed aperture size, the gain varies 10~dBi to 17~dBi in the designed broad frequency range.

To perform far-field measurements, the antenna distance should satisfy $R>2 D^{2} / \lambda$, where $\lambda$ is the wavelength, and $D$ is the antenna aperture size. The aperture diameter of the antenna under test (AUT) is 5.2~mm, and the shortest wavelength corresponding to 170~GHz is 1.76~mm. By calculation, the far-field distance must be greater than 30.7~mm. Our setup allows for a wide range of far-field distances for adjustment. For ease of installation and to ensure accurate measurement, we set the far-field distance to 1050~mm.

In the far-field measurements, the AUT is rotated around the axis located at about 1~mm away from the aperture towards to the waveguide, as the phase center of this design varies within 0.5-1.7~mm away from the aperture towards to the waveguide calculated from 80-170~GHz within $\pm$20~° in CST. The AUT is scanned from -90~° to 90~° with 1~° step. The cross-polarization is scanned at $\phi$=45~°, where the cross-polarization has the highest level. All far-field patterns are with the Ludwig’s 3rd definition.

\begin{figure}[h]
    \centering
    \begin{subfigure}{0.4\textwidth}
        \centering
        \includegraphics[width=\linewidth]{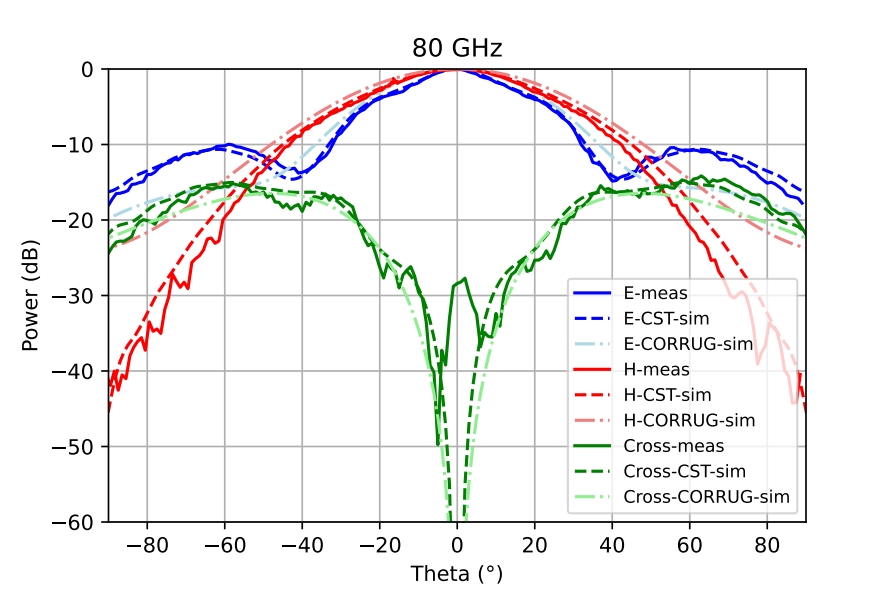}
    \end{subfigure}
    \begin{subfigure}{0.4\textwidth}
        \centering
        \includegraphics[width=\linewidth]{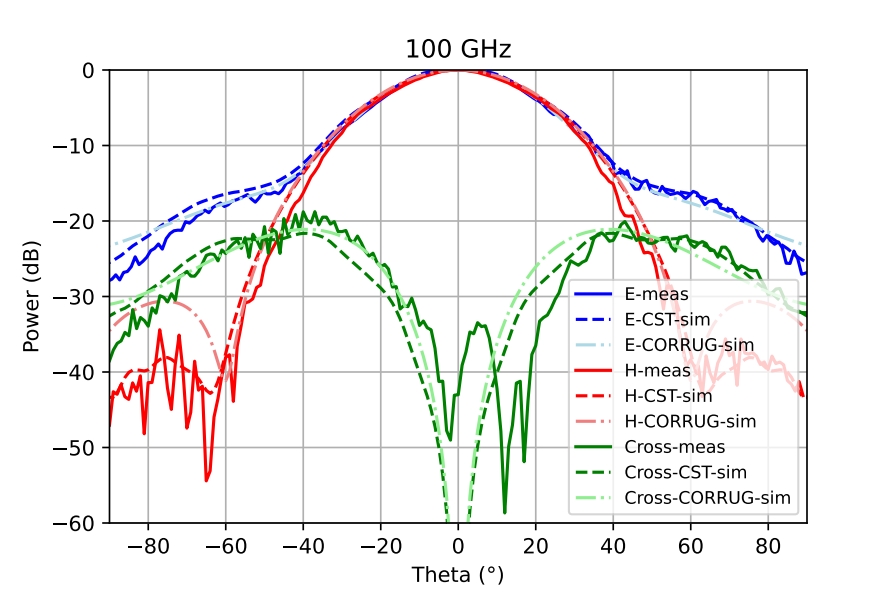}
    \end{subfigure}
    \begin{subfigure}{0.4\textwidth}
        \centering
        \includegraphics[width=\linewidth]{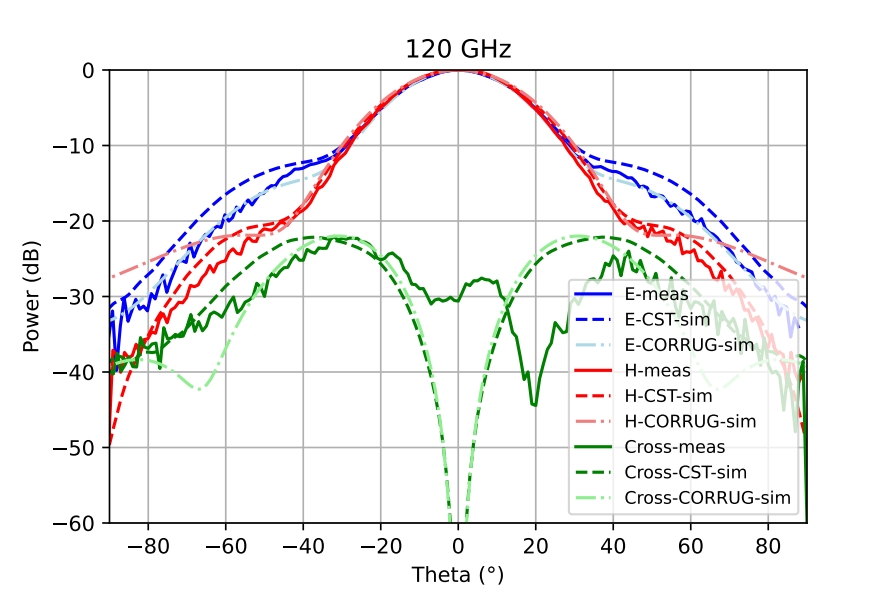}
    \end{subfigure}
    \begin{subfigure}{0.4\textwidth}
        \centering
        \includegraphics[width=\linewidth]{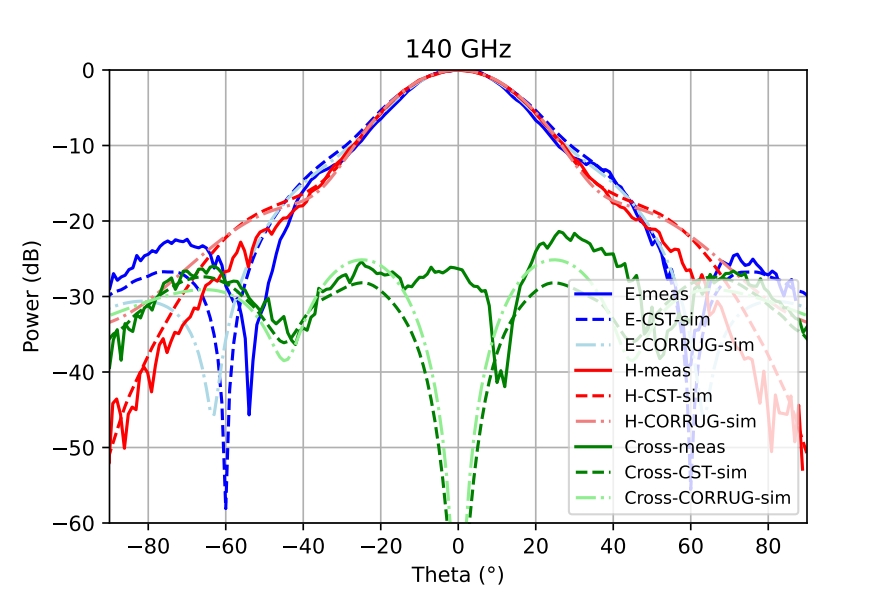}
    \end{subfigure}
    \begin{subfigure}{0.4\textwidth}
        \centering
        \includegraphics[width=\linewidth]{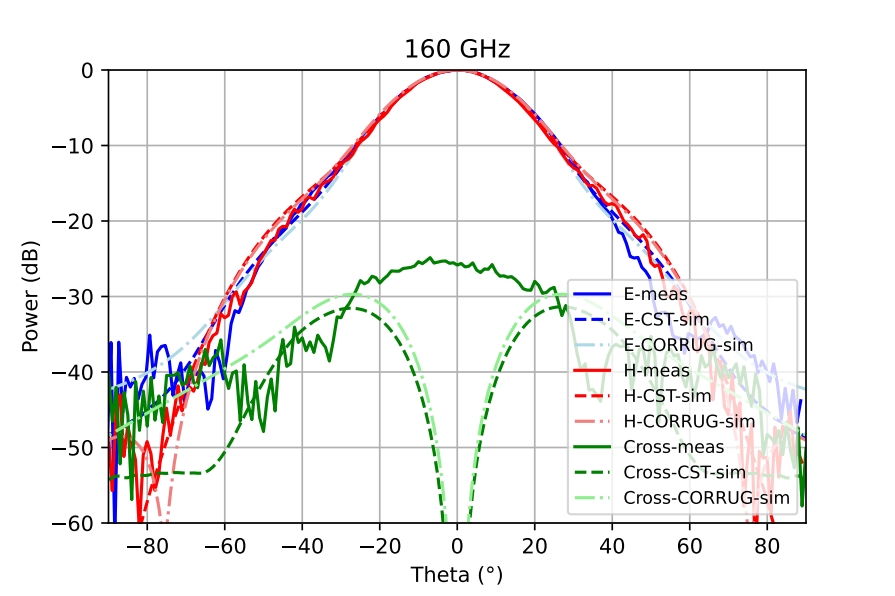}
    \end{subfigure}
    \caption{Far-field beam patterns of a fabricated silicon-plated horn.}
    \label{fig:beam1}
\end{figure}

Measurement results are shown in Fig.~\ref{fig:beam1} compared with the beams simulated by CORRUG software and CST software. The measurements are consistent with the CST results very well almost at all frequencies and angles. The CORRUG results agree with measurements when the angle is smaller than 40~°. This is mainly due to the mismatch between the aperture and the free-space in mode-matching calculations, as discussed in Sec.~\ref{sec:ant_opt}. As expected, when the aperture is 2.5 times larger than the wavelength, which corresponds to 144~GHz in our case, the measurements agree well with the CORRUG simulations. 

\begin{figure}[h]
    \centering
    \begin{subfigure}{0.4\textwidth}
        \centering
        \includegraphics[width=\linewidth]{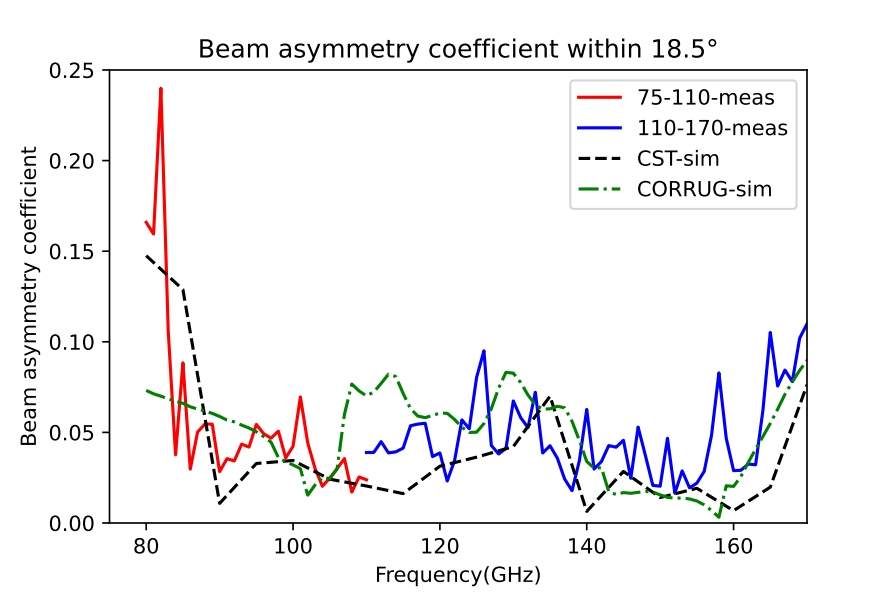}
        \tikz{(a) Beam asymmetry coefficient}
    \end{subfigure}
    \hspace{0.05\textwidth}
    \begin{subfigure}{0.4\textwidth}
        \centering
        \includegraphics[width=\linewidth]{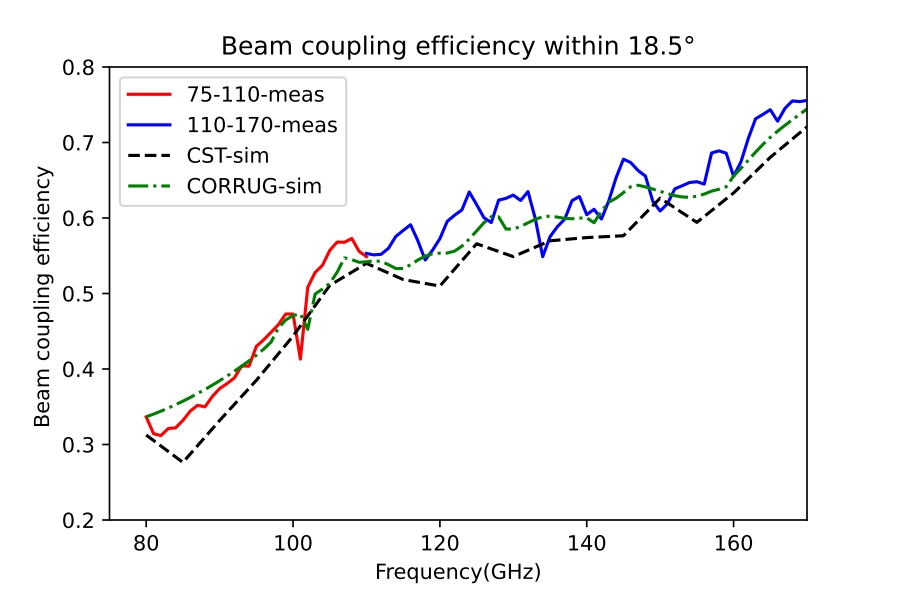}
        \tikz{(b) Beam coupling efficiency}
    \end{subfigure}
    \begin{subfigure}{0.4\textwidth}
        \centering
        \includegraphics[width=\linewidth]{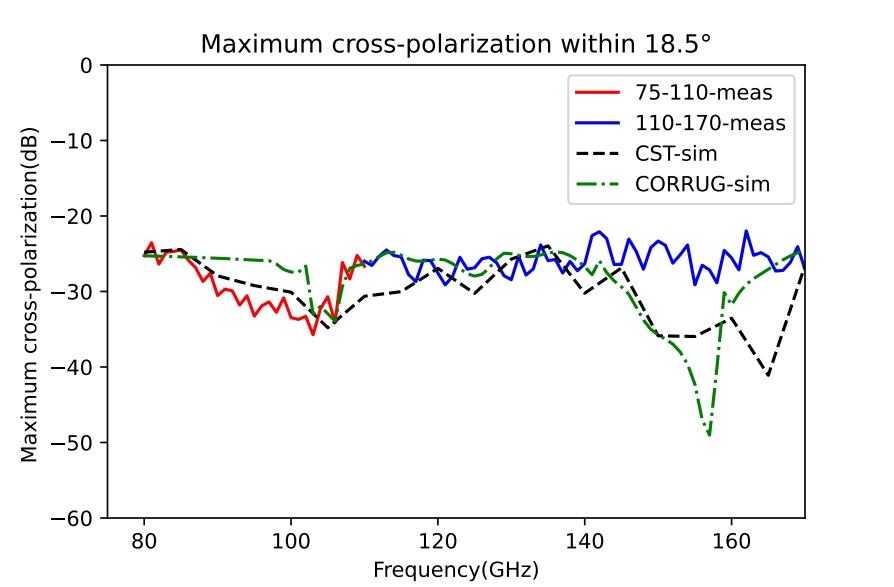}
        \tikz{(c) Maximum cross-polarization}
    \end{subfigure}
    \caption{Comparisons between the simulated results and the measured results of (a) the beam asymmetry coefficient. (b) the beam coupling efficiency, and (c) the cross-polarization.}
    \label{fig:result}
\end{figure}

The beam asymmetry coefficients, shown in Fig.~\ref{fig:result}, in general follow both simulations. Around 80~GHz, the asymmetry coefficient has a peak at 24~\%, which will be discussed in Sec.~\ref{sec:ant_array}. As the final beams will be the weighted beam for two frequency bands: 80-110~GHz and 125-165~GHz, the final beam asymmetry would still be smaller than 10~\%, the aim for our design. The beam coupling efficiency is well consistent with simulations. The maximum cross-polarizations within the cold stop are all smaller than -20~dB at all frequencies. 

\section{A 6-inch silicon-plated horn antenna array}
\label{sec:ant_array}

As the performance of a single silicon horn antenna is verified, the same design is applied to a 6-inch large array design for our 6-inch CMB detector module. The pixel design is briefly discussed in Ref.(~\cite{chai2024thermal}). Either transition edge sensors(~\cite{xu2024design}) or kinetic inductance detector(~\cite{chai2024thermal}) could be used. This new module design has 456 pixels in total, 7\% more than the detector module used by Simons Observatory(~\cite{mccarrick2021simons}), with a pitch size of 5.3~mm. 

\begin{figure*}
   \centering
   \includegraphics[width=0.97\textwidth]{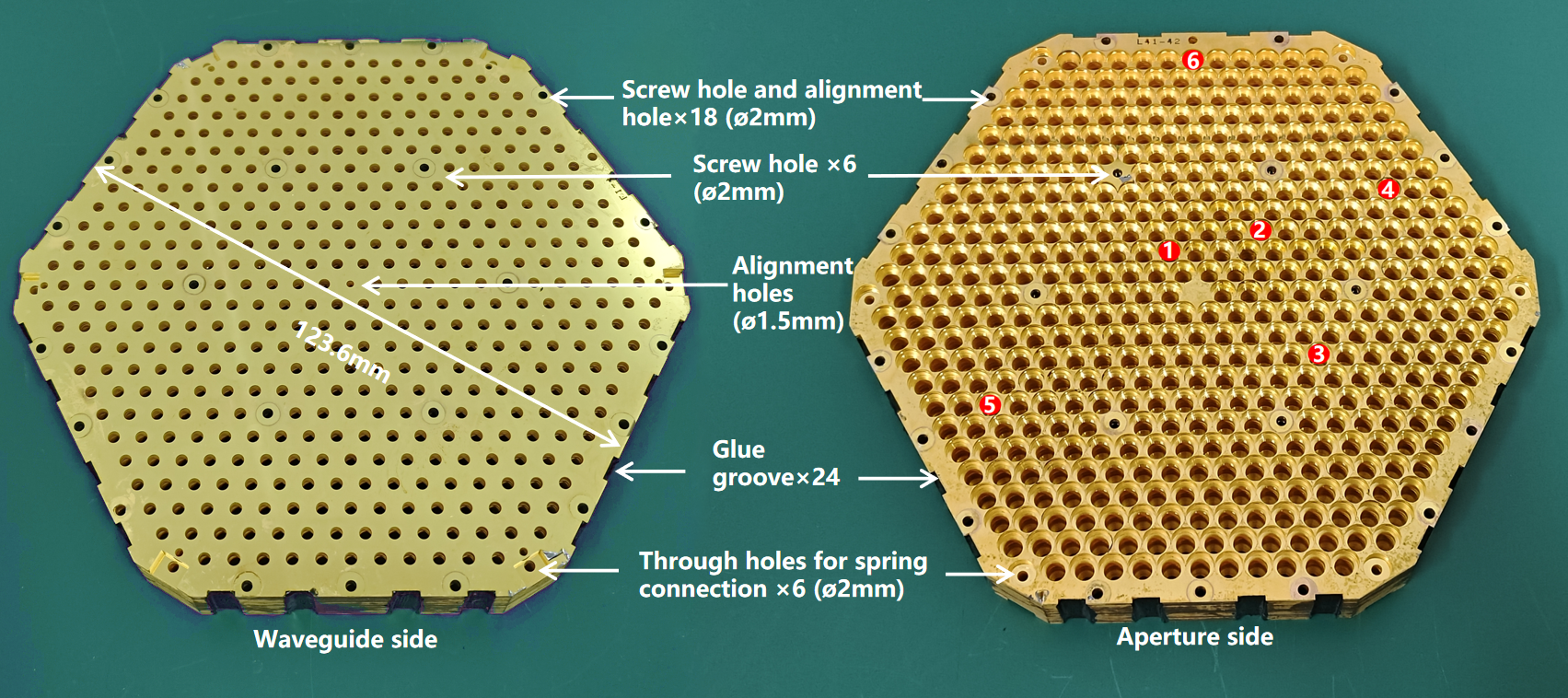}
   \caption{The front and back sides of the fabricated and assembled 6-inch large antenna array. Antenna labeled with numbers are selected for measurements.}
\label{fig:antenna module hole}
\end{figure*}

As the horn aperture is 5.2~mm, the narrowest width between horn apertures is only 100~$\mu\text{m}$. The total 51 layers in the single horn design, shown in Fig.~\ref{fig:ant profile}, are decreased to 42 layers by removing 4.5~mm-long circular waveguides at the beginning of the horn profile without changing the performance. The total length is decreased to 15~mm with 24 layers of 0.25~mm-thick 6-inch silicon wafers and 18 layers of 0.5~mm-thick 6-inch silicon wafers. The final assembled array is shown in Fig.~\ref{fig:antenna module hole}. In the ideal case, 469 antennas could be located, but 13 of them are used for packaging purposes. The central one of the array is revised to be the alignment hole for all detector related wafers. To fasten the array in gold plating, six of them in the middle of the array are used as through holes for screws. The six corners are also used as screw holes for the spring connections between the silicon antenna array and the metal packages to compensate the contraction at low temperature. On all six edges of the array, three screw holes and four glue grooves are designed for fastening.

\begin{figure}[h]
   \centering
   \includegraphics[width=0.5\textwidth,trim={5cm 3cm 0cm 0cm},clip]{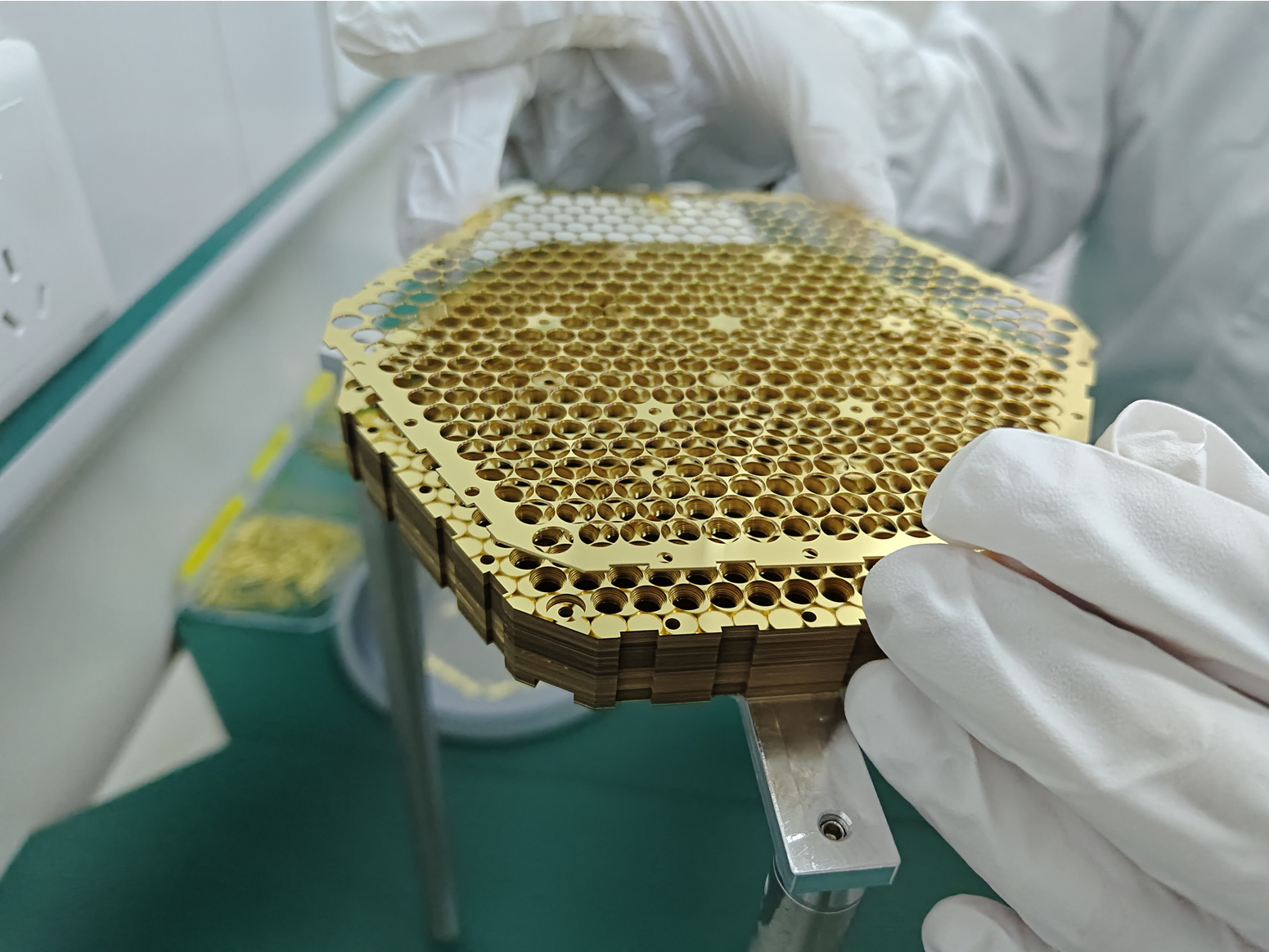}
   \caption{The process of stacking wafers of a 6-inch horn antenna array using a custom jig.}
\label{fig:antenna module assembly}
\end{figure}

The fabrication process is adapted from the single horn antenna fabrication process, shown in Fig.~\ref{fig:fab}. The full 6-inch wafer stacks dramatically increased the fabrication difficulty. After four-month fabrication in the public SINANO cleanroom, all wafers are fabricated with no noticeable broken structure. As shown in Fig.~\ref{fig:antenna module assembly}, the assembly was done on a custom jig in our cleanroom to prevent trapping particles between wafers. Wafers are aligned by two alignment pins with a tolerance of $\pm 1~\mu\text{m}$. After stacking, screws are used to secure the whole array for gold plating. A 2~$\mu\text{m}$-thick gold layer is plated as the antenna definition, and also to cover any possible gaps between wafers. Stycast 2850, which has a good thermal conductivity at low temperature, is used as the glue on the grooves of the module. The final antenna array is shown in Fig.~\ref{fig:antenna module hole}. Six labeled horn antennas are measured using a custom test tray. These antennas are positioned progressively farther from the module center to evaluate the uniformity and performance across the entire antenna array.

\begin{figure}[h]
   \centering
   \includegraphics[width=0.9\textwidth]{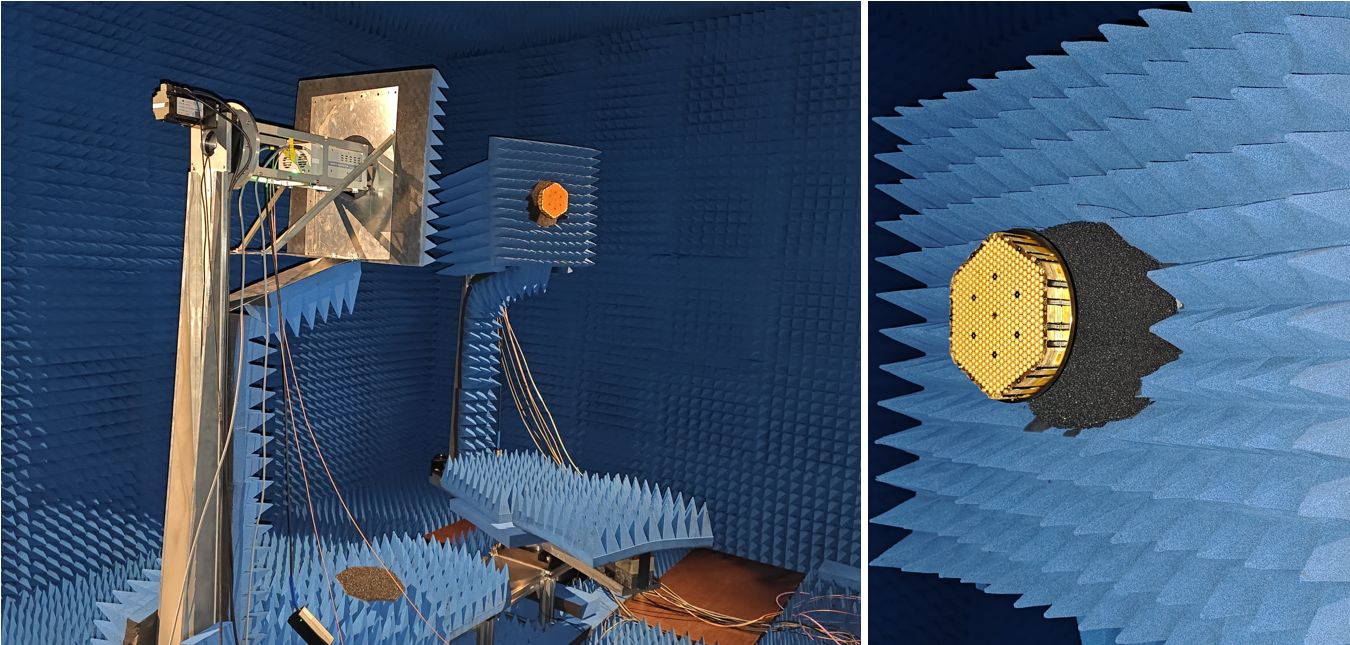}
   \caption{The far-field measurements setup in the IHEP microwave darkroom.}
\label{fig:array_meas}
\end{figure}

\begin{figure}[h]
    \centering
    \begin{subfigure}{0.4\textwidth}
        \centering
        \includegraphics[width=\linewidth]{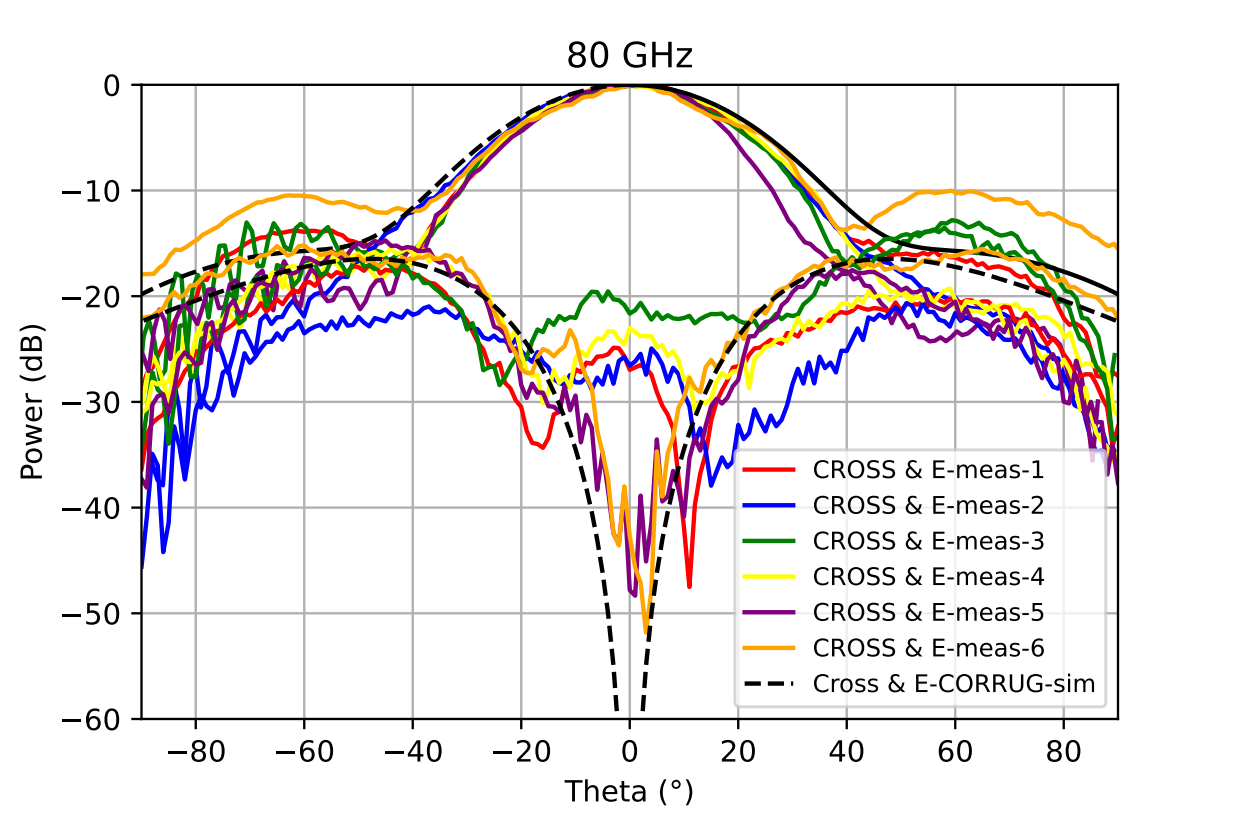}
    \end{subfigure}
    \begin{subfigure}{0.4\textwidth}
        \centering
        \includegraphics[width=\linewidth]{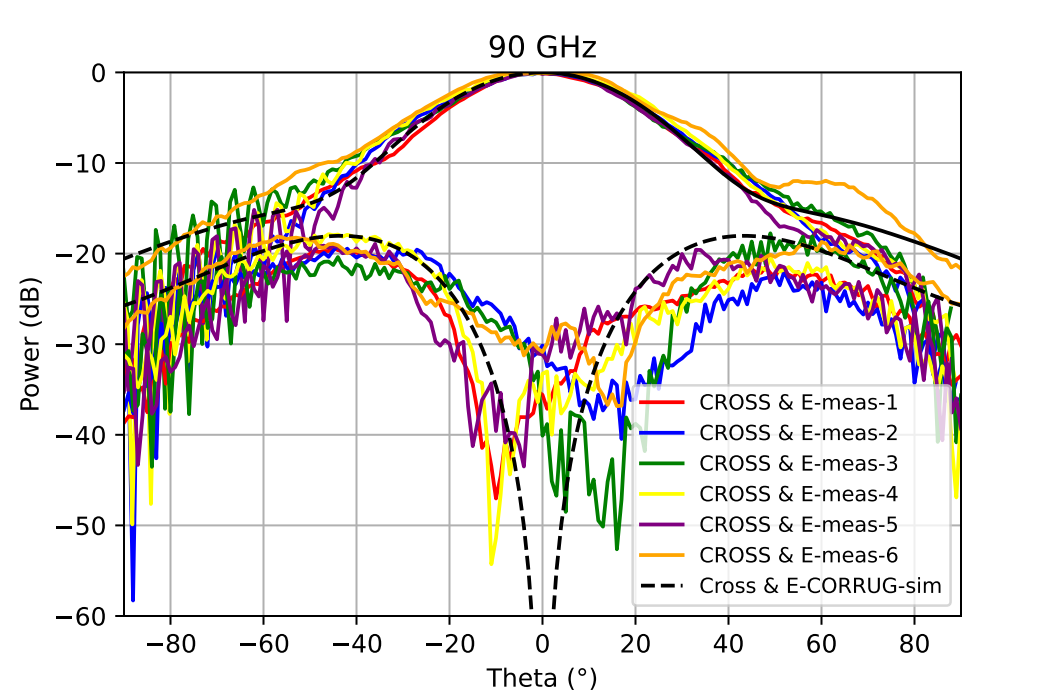}
    \end{subfigure}
    \begin{subfigure}{0.4\textwidth}
        \centering
        \includegraphics[width=\linewidth]{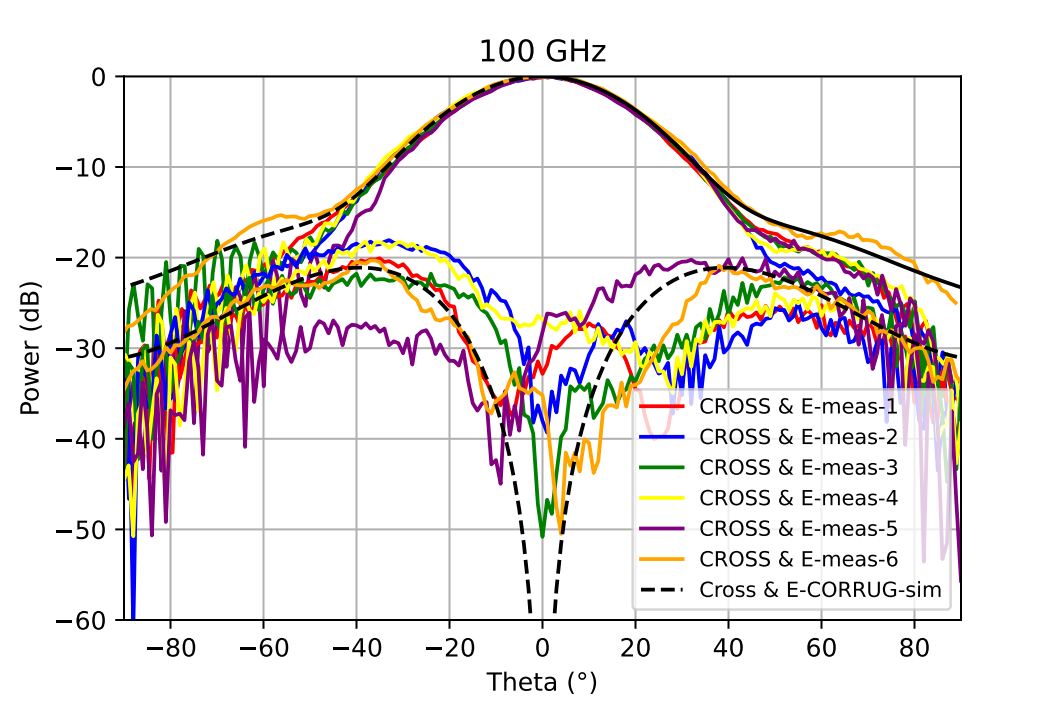}
    \end{subfigure}
    \begin{subfigure}{0.4\textwidth}
        \centering
        \includegraphics[width=\linewidth]{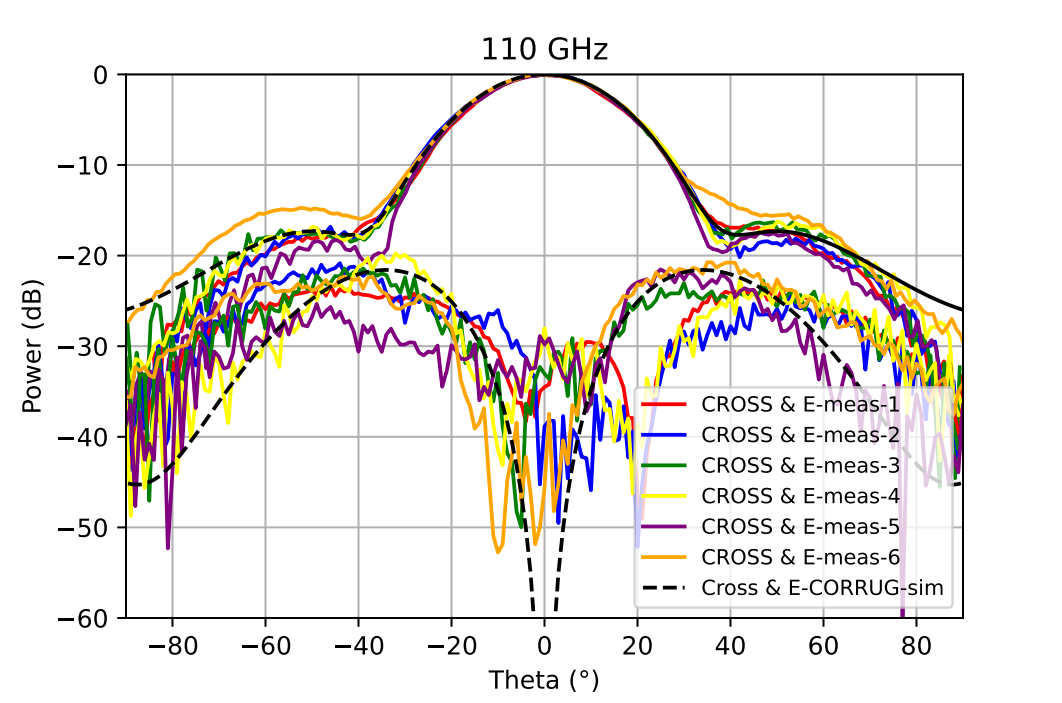}
    \end{subfigure}
    \begin{subfigure}{0.4\textwidth}
        \centering
        \includegraphics[width=\linewidth]{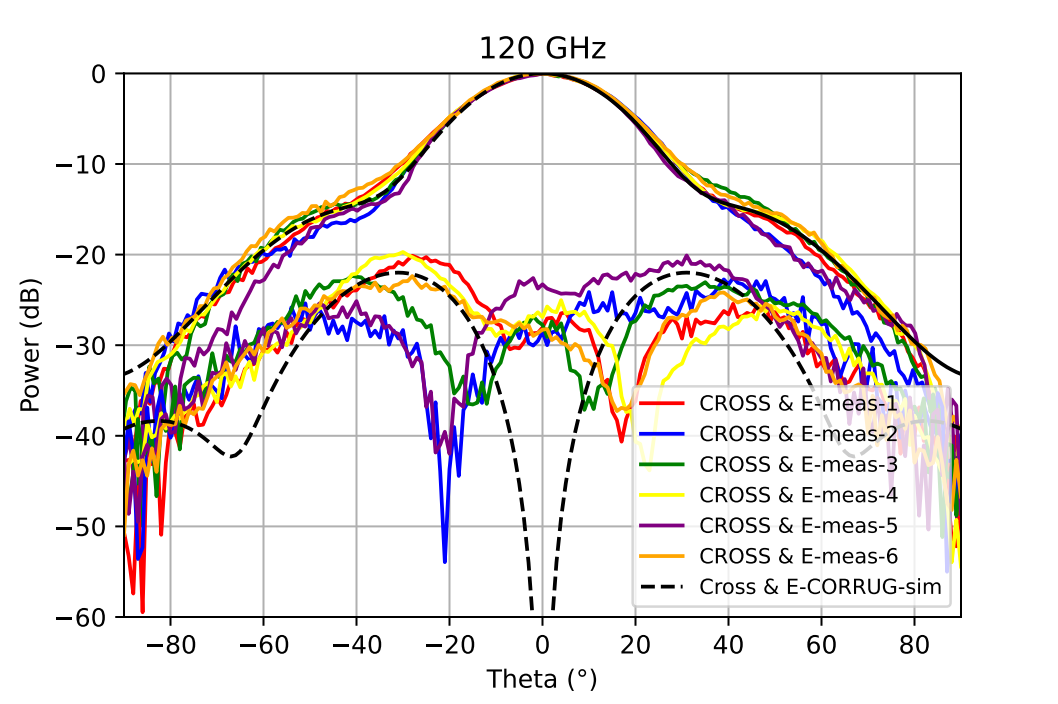}
    \end{subfigure}
    \begin{subfigure}{0.4\textwidth}
        \centering
        \includegraphics[width=\linewidth]{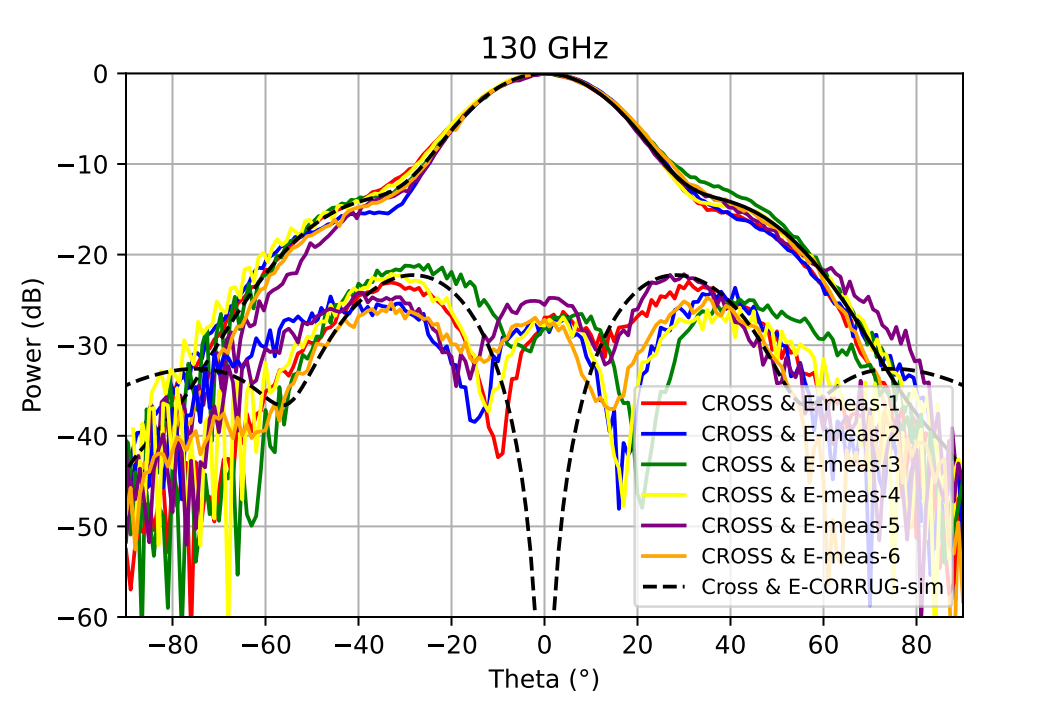}
    \end{subfigure}
    \begin{subfigure}{0.4\textwidth}
        \centering
        \includegraphics[width=\linewidth]{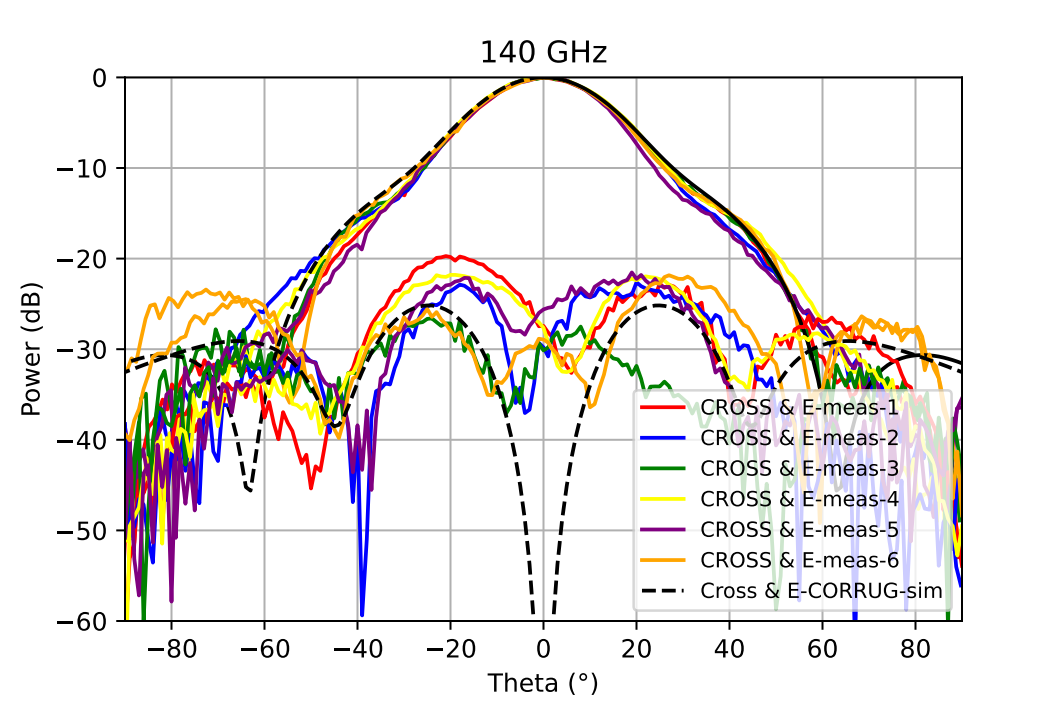}
    \end{subfigure}
    \begin{subfigure}{0.4\textwidth}
        \centering
        \includegraphics[width=\linewidth]{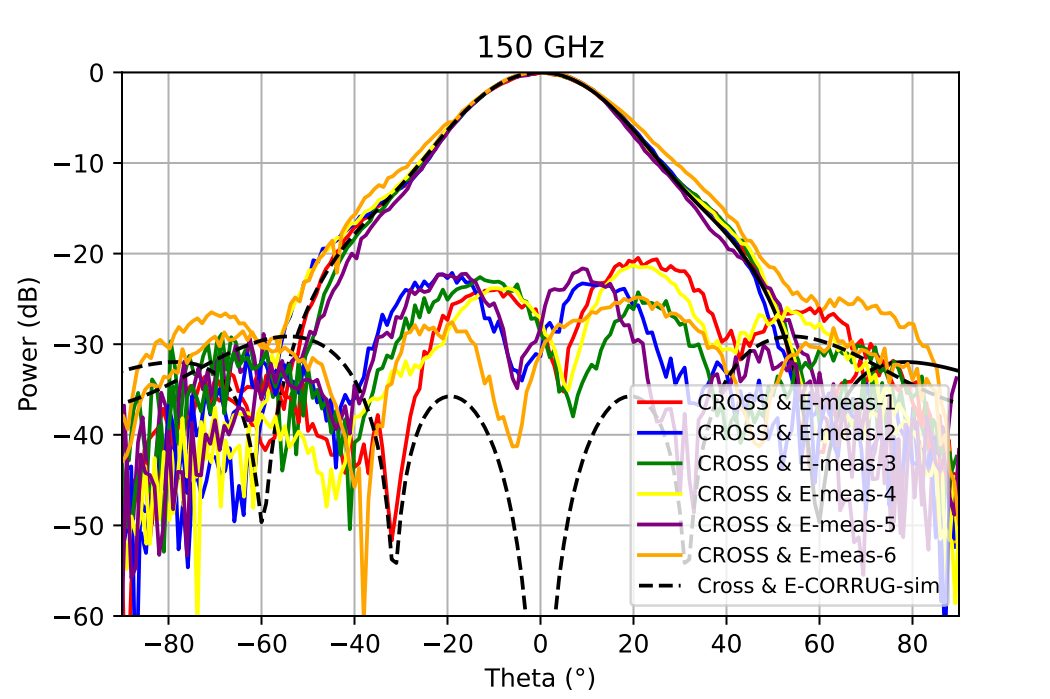}
    \end{subfigure}
    \begin{subfigure}{0.4\textwidth}
        \centering
        \includegraphics[width=\linewidth]{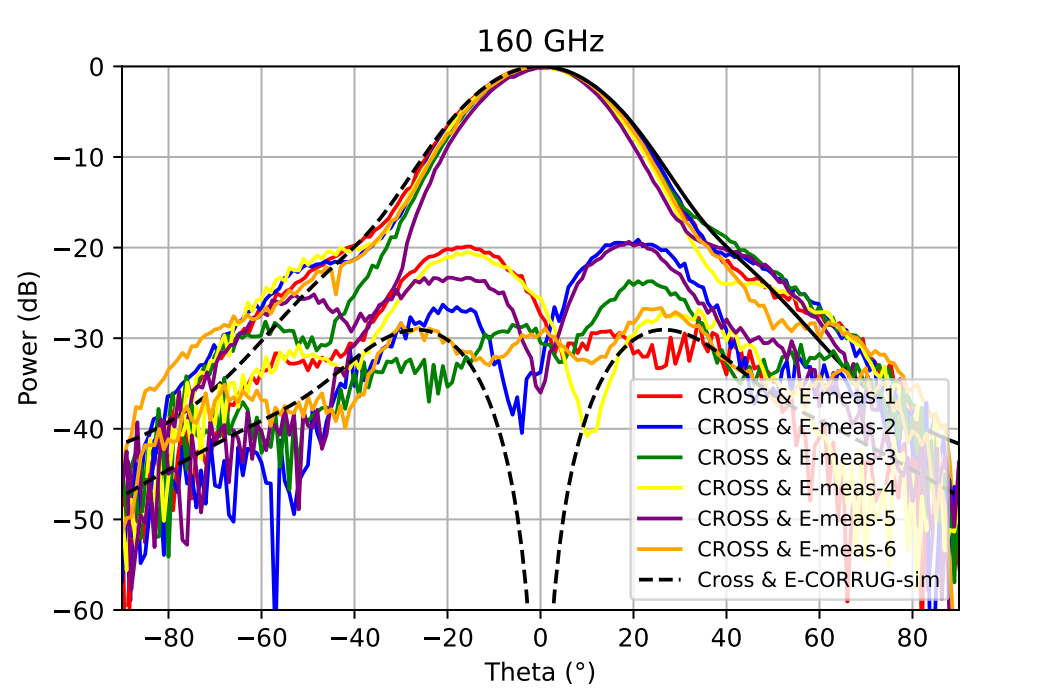}
    \end{subfigure}
    \begin{subfigure}{0.4\textwidth}
        \centering
        \includegraphics[width=\linewidth]{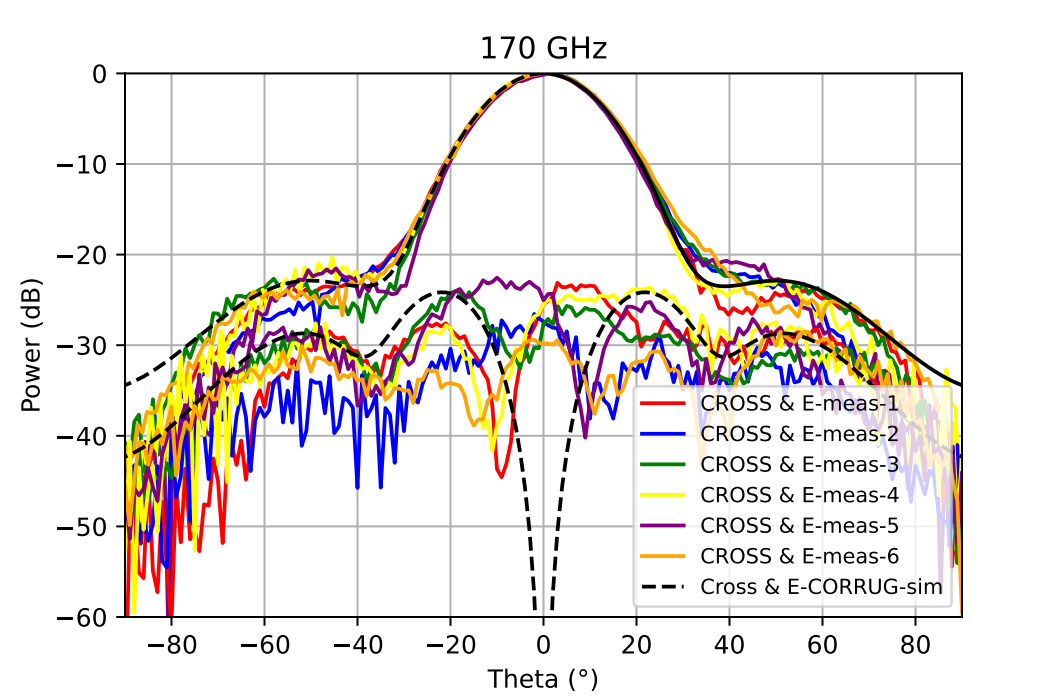}
    \end{subfigure}
    \caption{Measured E-plane and cross-polarization far-field beams of the 6 selected horns from 80~GHz to 170~GHz.}
    \label{fig:array E1}
\end{figure}

\begin{figure}[h]
    \centering
    \begin{subfigure}{0.4\textwidth}
        \centering
        \includegraphics[width=\linewidth]{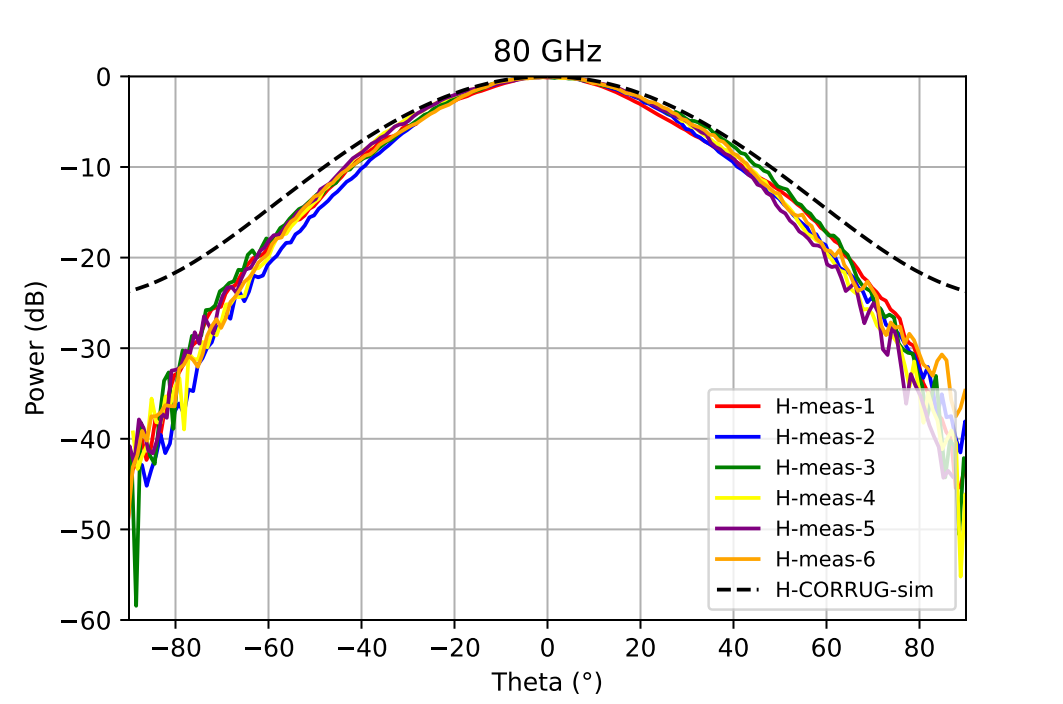}
    \end{subfigure}
    \begin{subfigure}{0.4\textwidth}
        \centering
        \includegraphics[width=\linewidth]{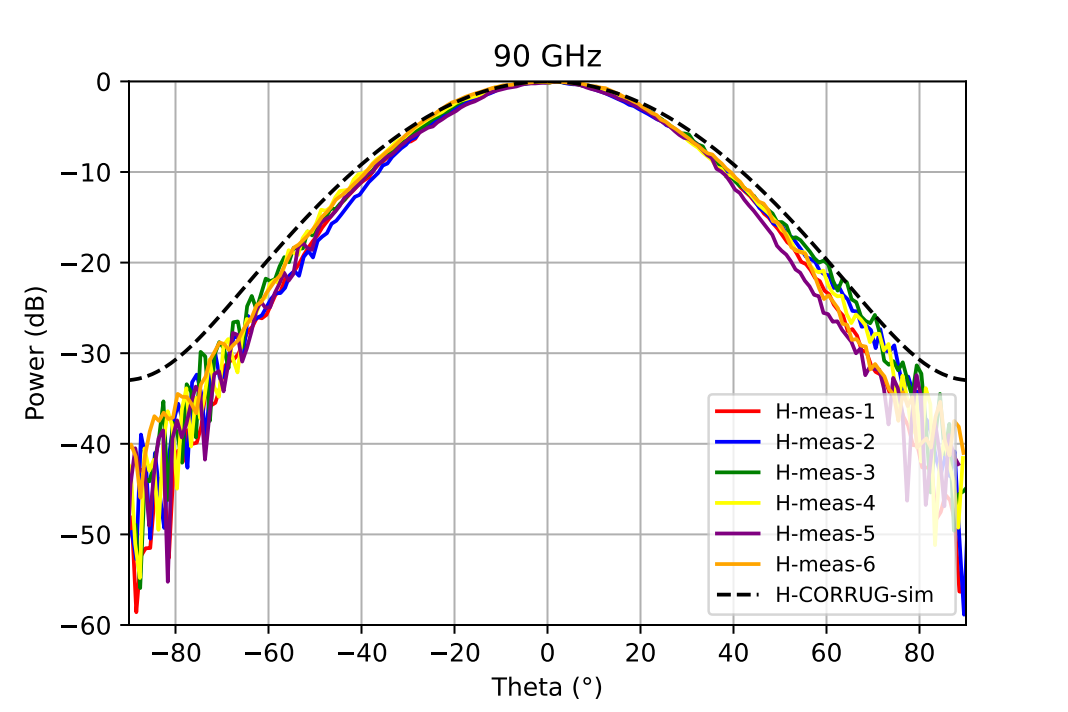}
    \end{subfigure}
    \begin{subfigure}{0.4\textwidth}
        \centering
        \includegraphics[width=\linewidth]{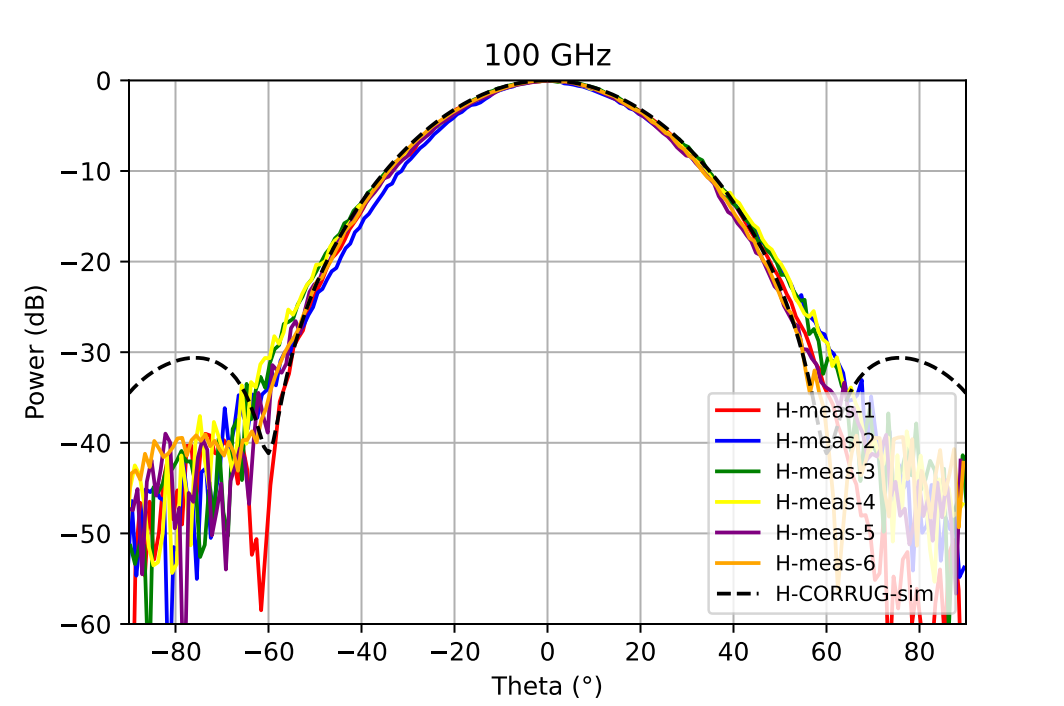}
    \end{subfigure}
    \begin{subfigure}{0.4\textwidth}
        \centering
        \includegraphics[width=\linewidth]{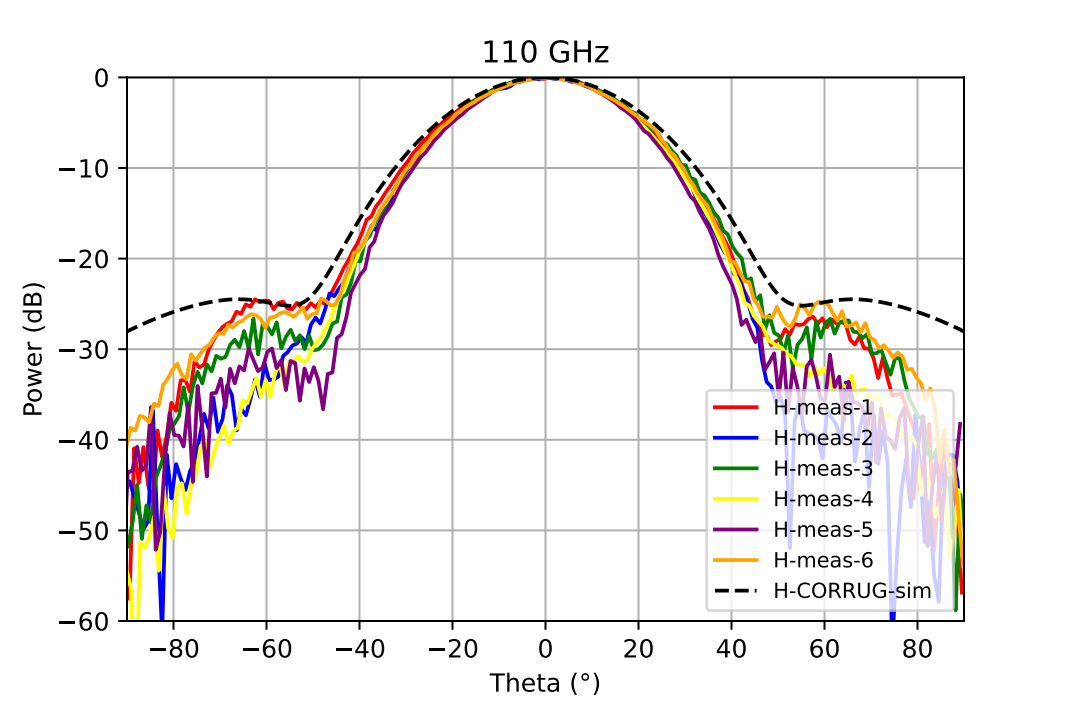}
    \end{subfigure}
    \begin{subfigure}{0.4\textwidth}
        \centering
        \includegraphics[width=\linewidth]{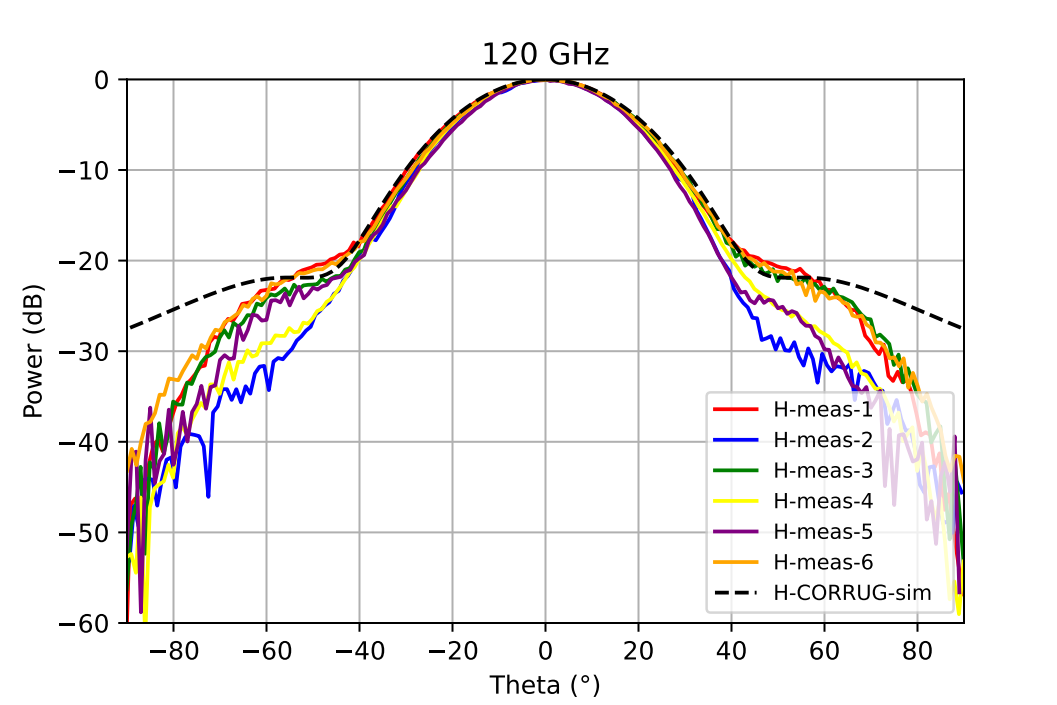}
    \end{subfigure}
    \begin{subfigure}{0.4\textwidth}
        \centering
        \includegraphics[width=\linewidth]{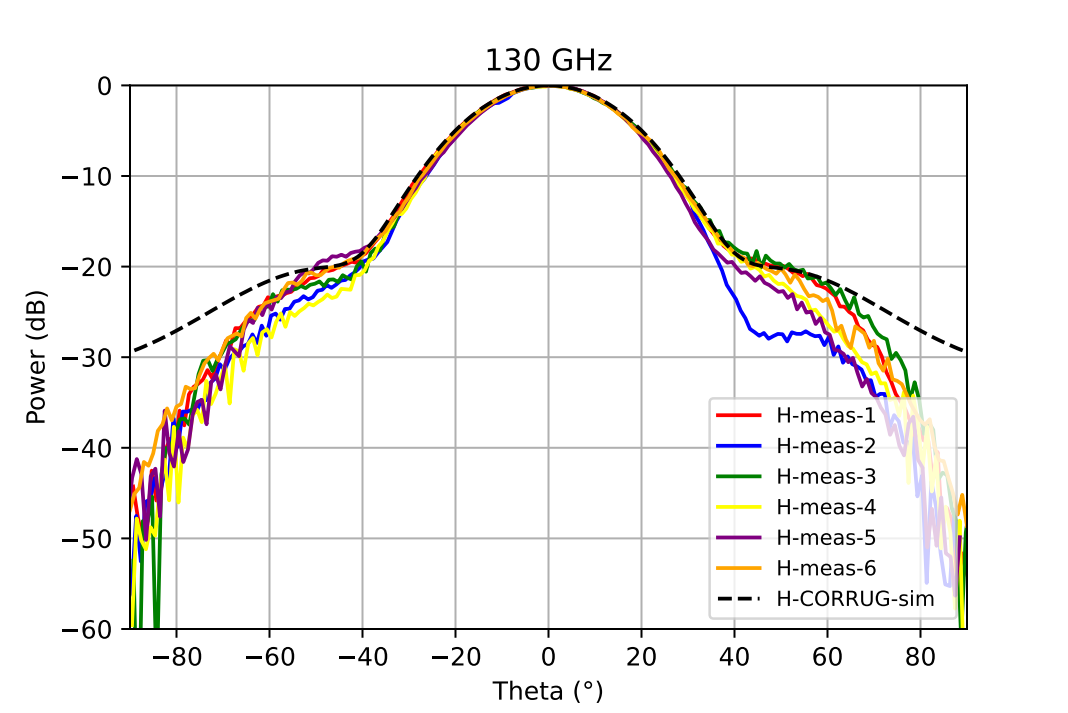}
    \end{subfigure}
    \begin{subfigure}{0.4\textwidth}
        \centering
        \includegraphics[width=\linewidth]{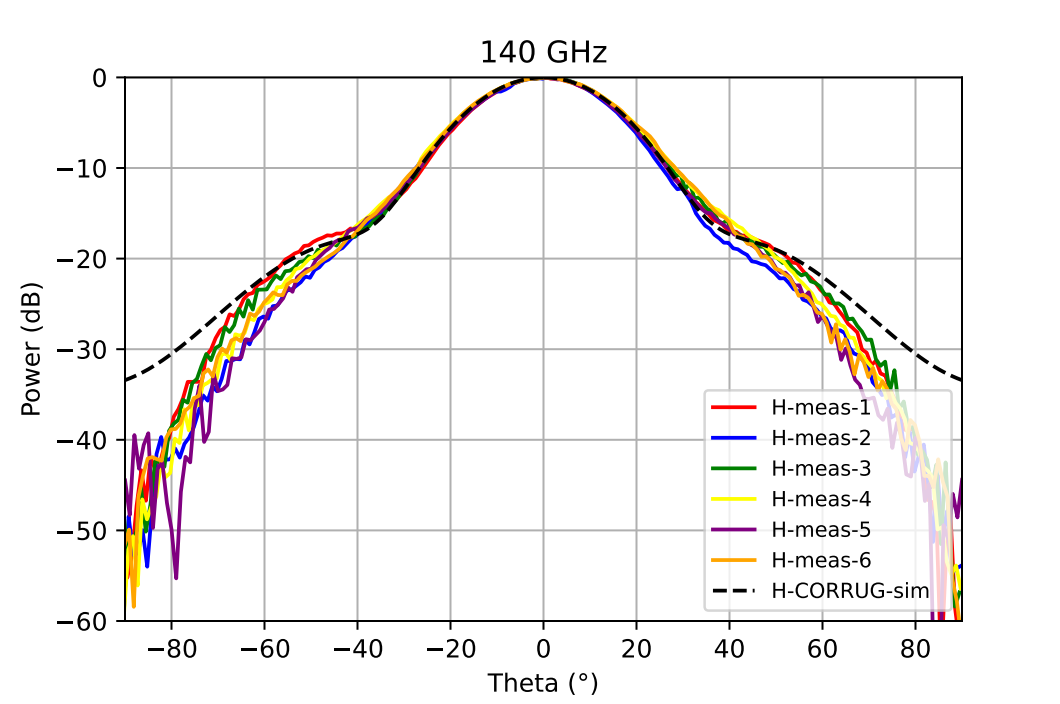}
    \end{subfigure}
    \begin{subfigure}{0.4\textwidth}
        \centering
        \includegraphics[width=\linewidth]{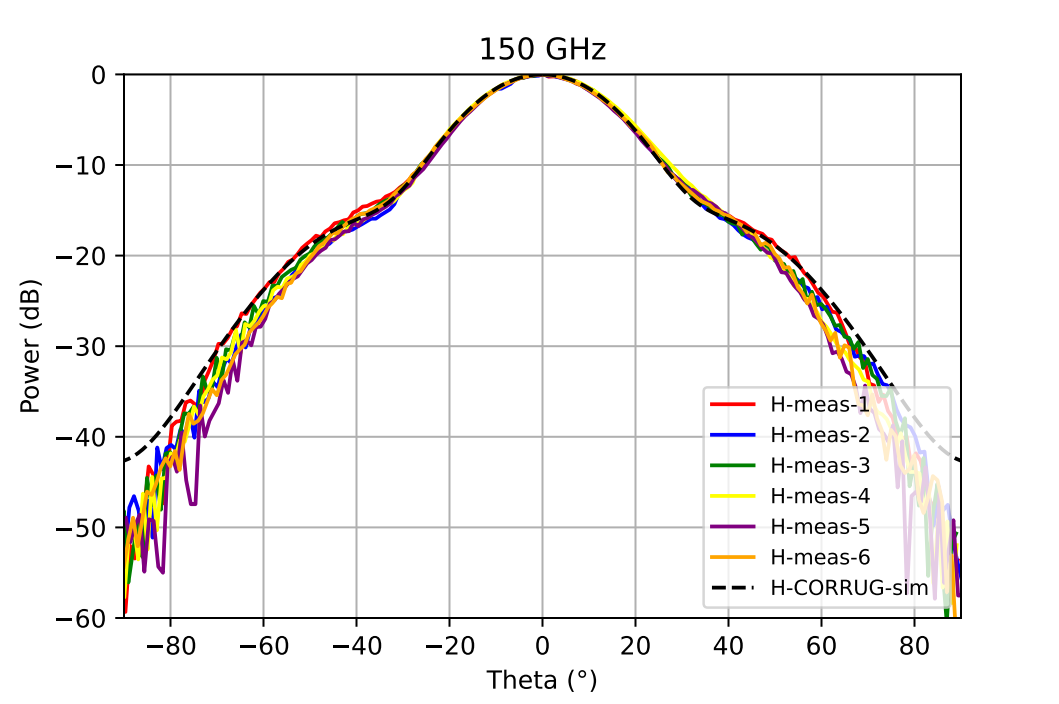}
    \end{subfigure}
    \begin{subfigure}{0.4\textwidth}
        \centering
        \includegraphics[width=\linewidth]{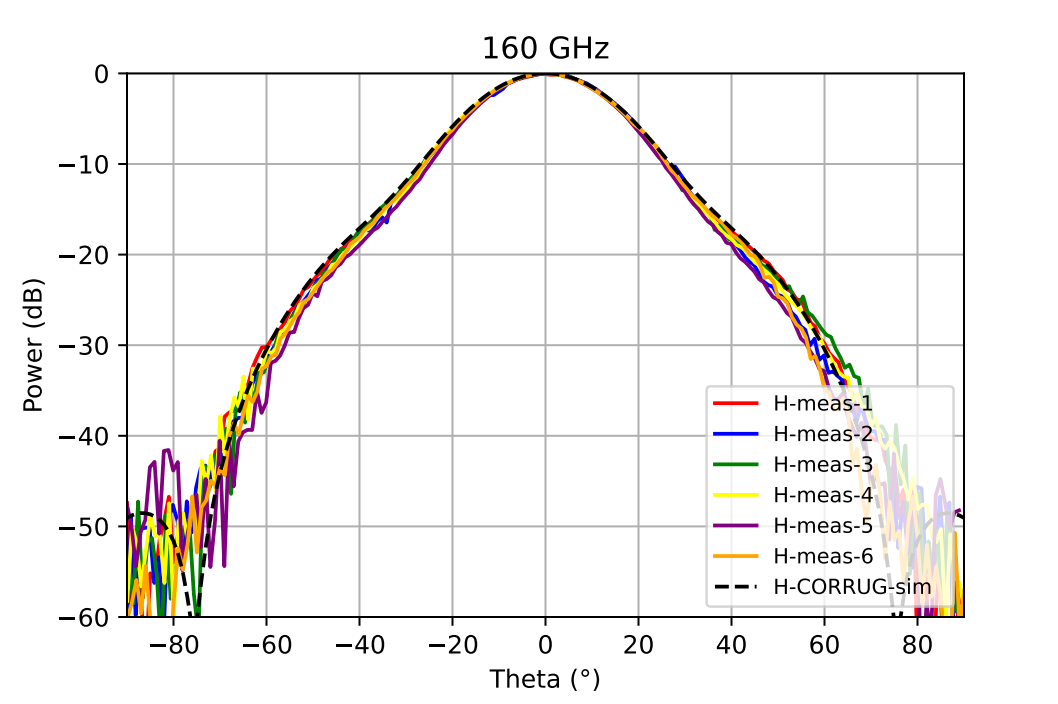}
    \end{subfigure}
    \begin{subfigure}{0.4\textwidth}
        \centering
        \includegraphics[width=\linewidth]{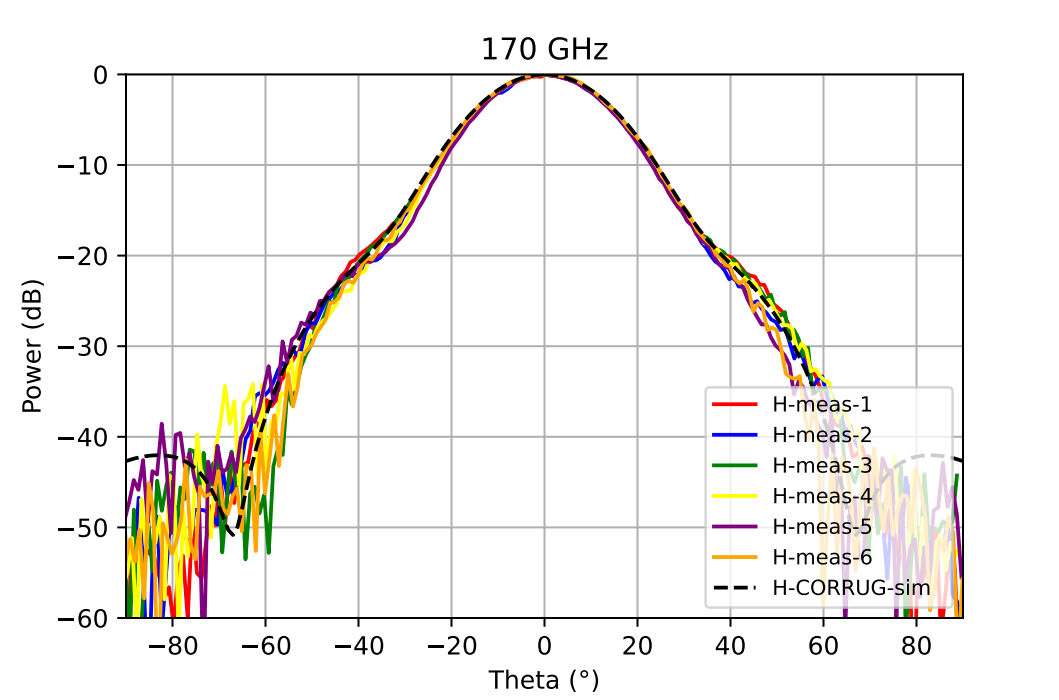}
    \end{subfigure}
    \caption{Measured H-plane far-field beams of the 6 selected horns from 80~GHz to 170~GHz.}
    \label{fig:array H1}
\end{figure}

\begin{figure}[h]
    \centering
    \begin{subfigure}{0.4\textwidth}
        \centering
        \includegraphics[width=\linewidth]{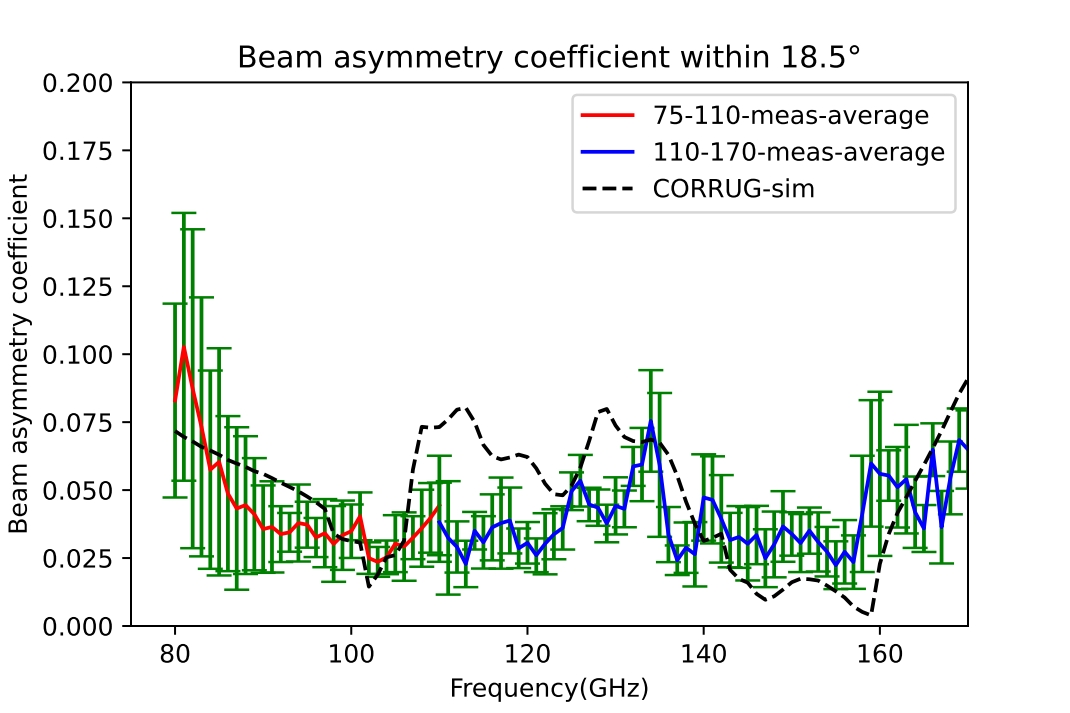}
        \tikz{(a) Beam asymmetry coefficient}
    \end{subfigure}
    \hspace{0.05\textwidth}
    \begin{subfigure}{0.4\textwidth}
        \centering
        \includegraphics[width=\linewidth]{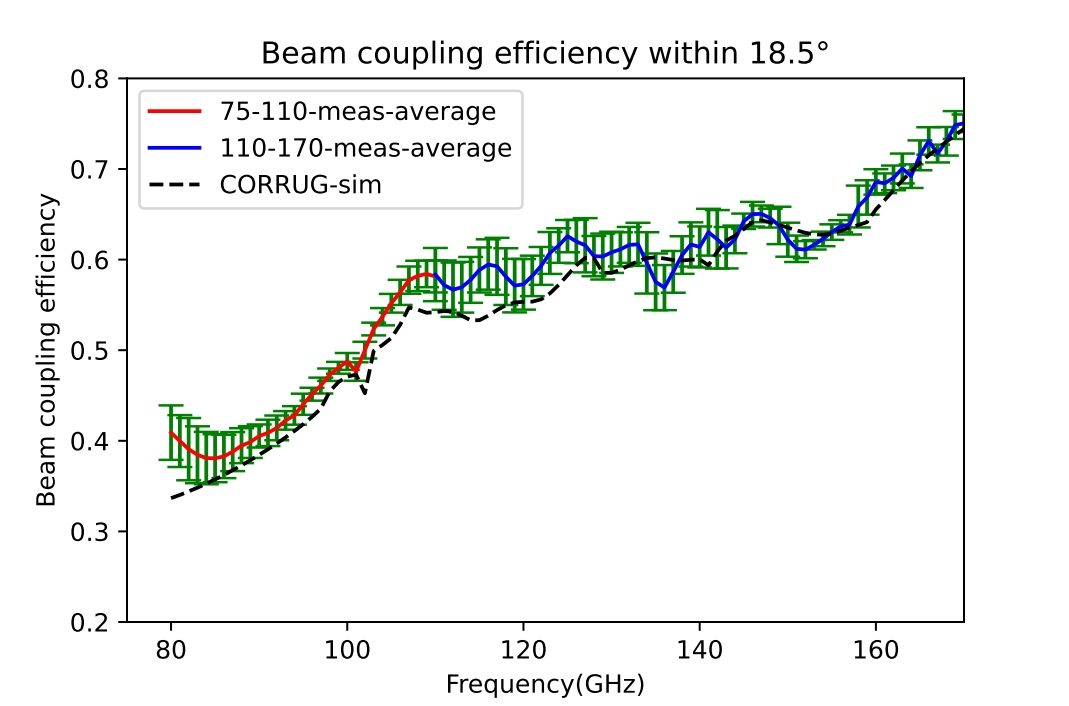}
        \tikz{(b) Beam coupling efficiency}
    \end{subfigure}
    \begin{subfigure}{0.4\textwidth}
        \centering
        \includegraphics[width=\linewidth]{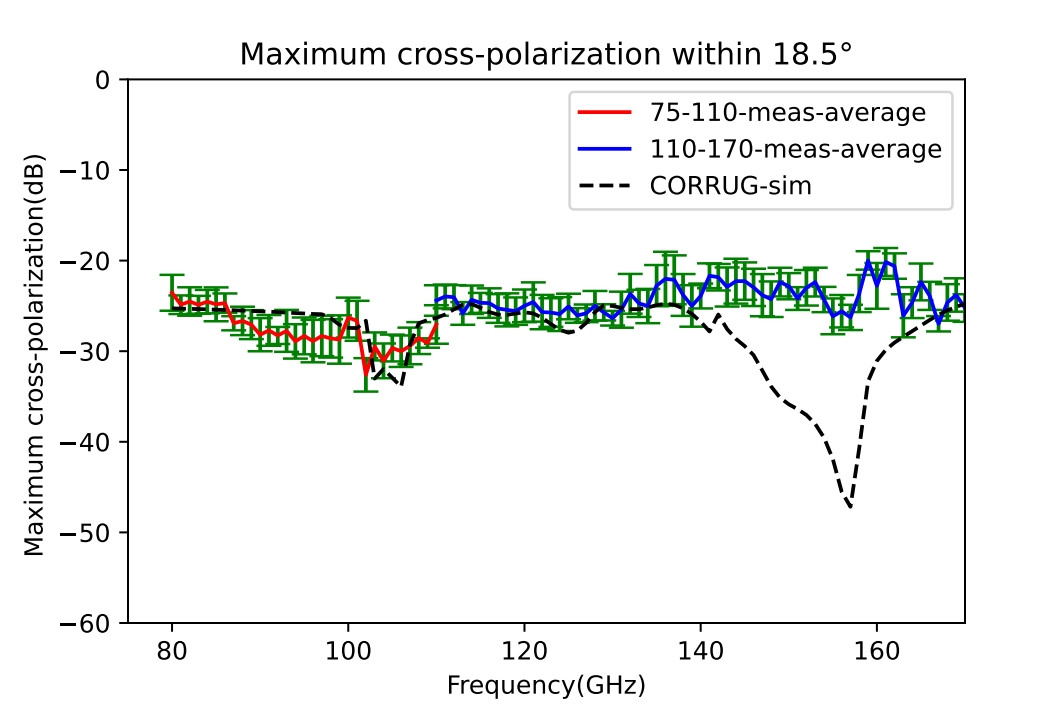}
        \tikz{(c) Maximum cross-polarization}
    \end{subfigure}
    \caption{Measurements of (a) beam asymmetry coefficient, (b) beam coupling efficiency, and (c) maximum cross-polarization obtained through far-field testing of the six horns.}
    \label{fig:result-array}
\end{figure}

The far-field setup, shown in Fig.~\ref{fig:array_meas}, is the same as the single antenna measurements performed in the IHEP darkroom. The beam patterns for the six horn antennas are shown in Fig~\ref{fig:array E1} and Fig~\ref{fig:array H1}. In general, the measurements are consistent with simulations, especially when the angle is within 20~°. When the angle is larger than 80~°, the covered absorber used to remove reflections in measurements begins attenuating certain beam power giving the measured beam lower than simulations. All side lobes are below -10~dB as expected. Cross polarizations are smaller than -20~dB for most of measured beams within 20~°. The average maximum cross polarization within 18.5~° is lower than -20~dB shown in Fig~\ref{fig:result-array}.

The average beam asymmetry coefficient is smaller than 10\% at all frequencies except 82~GHz, consistent with simulations. For frequencies smaller than 115~GHz, the beam asymmetry has an unexpected large variation compared with our fabrication error analysis in Fig.~\ref{fig:combine}. In this error analysis the alignment errors were not considered, as the misalignment of two circles with similar radii gives an irregular overlap resulting in a calculation difficulty. Hu et al.(~\cite{hu2021design}) has simulated the alignment errors for silicon-plated corrugated horns. Their statical results show that the beam waists at frequencies smaller than $1.3f_{\textrm{cut-off}}$ have a larger standard deviation than high frequencies results. In our case, $f_{\textrm{cut-off}}=76.4$~GHz and from 100~GHz ($1.3f_{\textrm{cut-off}}$) the scattering of the beam asymmetry becomes much smaller consistent with Ref.(~\cite{hu2021design}). We suggest that the wave propagation parameters vary fast around $f_{\textrm{cut-off}}$ and even a small change in $f_{\textrm{cut-off}}$ caused by the misalignment will change beam dramatically. Although the final in-band (80-110~GHz and 125-165~GHz) beam asymmetry of this array would still be smaller than 10~\% suitable for CMB observations, this seems to be a typical problem for silicon plated horn antennas. Metal horn arrays may not have this problem, if the final profile is machined by the same tool.

The coupling efficiency is consistent with simulation. As the H-plane is lower than simulation at large angles and the absorber attenuates some signal, the coupling efficiency is a little higher than simulation below 130~GHz. The high uncertainty around 80~GHz is also due to the aforementioned misalignment.  

So far, the performance of the presented antenna arrays are capable for observations. For future possible improvements, we may improve the misalignment by applying more alignment pins and using even thicker wafers to decrease the number of layers. New algorithm for simulating non-circular waveguide would also help the error misalignment analysis. Also, with current 1.54~F$\lambda$ aperture size, we have shown that the optimizations could be improved by combing 3D simulations as a final together with mode-matching simulations, especially for low frequencies beams. We may also try lowering the cut-off frequency for a better beam asymmetry, but the asymmetry at high frequency will be worse. Optimizations should be done to further explore the parameter space.

\section{Conclusion}
\label{sec:concl}

We have designed and fabricated 80-170~GHz broadband silicon-plated smooth-walled horn antenna arrays for primordial gravitational wave search. The arrays have 456 horn antennas based on 6-inch micro-fabrication process. Measurement results are consistent with simulations. The overall in-band cross polarization is smaller than -20~dB and the in-band beam asymmetry is smaller than 10\%. This antenna arrays will be packaged together with detector arrays as a focal plane module for future CMB projects like AliCPT. Possible improvements are also discussed.

\section{Acknowledgment}
\label{sec:ack}

The authors thank Dr. Jie Hu from Purple Mountain Observatory for useful discussion. This work is supported by National Key Research and Development Program of China (Grants No. 2022YFC2205000 and No. 2021YFC2203400).

\bibliographystyle{raa}

\bibliography{ref}

\end{document}